\documentclass[aps,pra,
reprint,
superscriptaddress,
 amsmath,amssymb,
floatfix,
]{revtex4-2}

\usepackage[caption = false]{subfig}
\usepackage{floatrow}
\usepackage{graphicx}
\usepackage{dcolumn}
\usepackage{bm}
\usepackage{braket}
\usepackage{hhline}
\usepackage{array}
\usepackage{booktabs}
\usepackage{nicefrac}
\usepackage{chemformula}
\usepackage{xspace}
\usepackage{tabularx}
\usepackage{textcomp} 

\newcommand{\HeFour}{\ch{^4 He}}
\newcommand{\HeThree}{\ch{^3 He}}
\newcommand{\hestar}{\ch{He}*}
\newcommand{\HeThreeStar}{\ch{^3 He}*} 
\newcommand{\HeFourStar}{\ch{^4 He}*}
\newcommand{\MetastableState}{2^{3\!}S_1}%
\newcommand{\CoolState}{2^{3\!}P_{2} }%

\usepackage{seqsplit}
\usepackage[hidelinks,colorlinks]{hyperref}
\hypersetup{citecolor=[rgb]{0.25,0.14,0.63}}
\hypersetup{urlcolor=[rgb]{0.25,0.14,0.63}}
\hypersetup{linkcolor=black}

\begin{document}

\preprint{APS/123-QED}

\author{Kieran F. Thomas}
\author{Zhuoxian Ou}
\author{Bryce M. Henson}
\author{Angela A. Baiju}
\author{Sean S. Hodgman}
\author{Andrew G. Truscott}

\affiliation{Department of Quantum Science and Technology, Research School of Physics, The Australian National University, Canberra, ACT 2601, Australia}

\title{Production of a highly degenerate Fermi gas of metastable helium-3 atoms}

\begin{abstract}
    We report on the achievement of quantum degeneracy in both components of a Bose-Fermi mixture of metastable helium atoms, \HeFourStar{} and \HeThreeStar{}.
    Degeneracy is achieved via Doppler cooling and forced evaporation for \HeFourStar{}, and sympathetically cooling \HeThreeStar{} with \HeFourStar{}. We discuss our simplified implementation, along with the high versatility of our system.
    This technique is able to produce a degenerate Fermi gas with a minimum reduced temperature of \(T/T_F=0.14(1)\), consisting of \(2.5 \times 10^4\) \HeThreeStar{} atoms. Due to the high internal energy of both isotopes single atom detection is possible, opening the possibility of a large number of experiments into Bose-Fermi mixtures.
\end{abstract}
\maketitle

\section{Introduction}

Since the first realization of Bose - Einstein condensation (BEC) in trapped atomic clouds in 1995 \cite{anderson1995observation}, BECs have provided a great range of experimental possibilities, such as atom optics \cite{Jeltes2007}, the study of the interaction between light and matter\cite{doi:10.1126/science.abk2502}, and quantum computation \cite{PhysRevA.85.040306,BYRNES2015102}. This is because they offer a great degree of control and ease of measurement in a large quantum system. Like bosons, ultracold fermions also offer insights into numerous quantum phenomena, for example high temperature superconductivity \cite{murthy2018high}, and fermionic superfluidity \cite{zwierlein2005vortices,chin2006evidence}, and have hence been an active field of experimental study in recent years \cite{RevModPhys.80.885,RevModPhys.80.1215}. Furthermore, degenerate bose-fermi mixtures have provided access to an even wider array of possible physics, such as phase separation \cite{PhysRevB.78.134517,PhysRevA.78.061606,PhysRevA.81.013622}, and polaron physics \cite{PhysRevA.84.011601,PhysRevA.103.053314}. While there have been a number of successful investigations into degenerate fermi gases (DFGs) and bose-fermi mixture \cite{doi:10.1126/science.285.5434.1703,doi:10.1126/science.1059318,Onofrio_2016,PhysRevLett.89.150403,PhysRevLett.96.180402,PhysRevLett.97.080404}, these investigations are generally less common compared to the pure bosonic counterparts. This is primarily due to the increased experimental complexity that fermionic gases bring, as the evaporative cooling that has allowed access to these ultracold temperatures is not directly possible for fermions. To circumvent this issue the fermionic species is often cooled to degeneracy sympathetically with a bosonic partner species \cite{PhysRevA.64.011402,PhysRevLett.97.080404}, as presented in this work.

Metastable helium (referring to \HeFourStar{} or \HeThreeStar{} in their respective \(2^3S_1\) excited electronic states) provides an especially unique platform for the study of dilute quantum gases due to its high internal energy, allowing for efficient single atom detection (for either isotope). This gives access to the quantum many-body wavefunction via measurements of momentum correlations in the time-of-flight profiles, and thus can be used to test many important questions in the field. For example, \HeFourStar{} has been used to observe atomic analogues of quantum optics phenomena such as Hanbury Brown-Twiss bunching \cite{Jeltes2007,Manning:10}, and wavelike interference in atomic systems \cite{Manning2015,Lopes2015,PhysRevLett.119.173202}, as well as precision tests of quantum electrodynamics \cite{vanRooij196,Rengelink2018,doi:10.1126/science.abk2502}.

Being able to detect individual atoms of a Bose - Fermi mixture in the far-field will also open the possibility of a range of experiments such as the interplay of the two statistics, the production and measurement of a mass entangled state for quantum tests of gravity \cite{PhysRevLett.120.043602}, and high precision spectroscopic measurements of each isotope allowing for precision test of quantum theory \cite{Rengelink2018}. In this paper, we report on the production of a degenerate Bose - Fermi mixture of bosonic helium (\HeFourStar{}) and fermionic helium (\HeThreeStar{}). Following laser cooling of both species, the bosons are first cooled evaporatively and then used as a sympathetic coolant for the fermions. The achievable temperatures relative to their respective critical temperatures and number of atoms are presented and analyzed. We describe our novel approach of modifying our existing apparatus for the cooling and trapping of \HeFourStar{} \cite{Dall2007} to accommodate \HeThreeStar{} with minimal alterations to the vacuum and laser system. This system is capable of demonstrating and studying many body correlations and could shed light on some of the most interesting problems in modern quantum physics.

\section{Background}
In this section we will cover both the theoretical basis of the cooling techniques applied to \HeThreeStar{} and \HeFourStar{}, highlighting the most pertinent distinctions between the two species (arising primarily from differing electronic structures and masses).

\subsection{Degeneracy Temperature for Bosonic and Fermionic matter} 

From their respective symmetries under particle exchange we find the behavior of bosons (symmetric) and fermions (antisymmetric) are governed by Bose-Einstein statistics and Fermi-Dirac statistics respectively. At high temperature, quantum statistics become negligible and both of these kinds of particles obey the classical Maxwell-Boltzmann distribution. However, for both fermions and bosons we can define a respective temperature below which quantum effects become apparent, and thus starkly different behavior is displayed both between one another and their classical counterparts.

For the bosons, below the condensation temperature $T_C$, a macroscopic fraction of the total number of particles will occupy the lowest-energy single-particle state, forming a Bose-Einstein condensate (BEC). If we have \(N_b\) bosons in a harmonic trap then the phase transition temperature \(T_C\) can be obtained as \cite{pethick2008bose}
\begin{align}
    k_B T_C = 0.94 \hspace{0.1 cm} \hbar \bar{\omega}_b N_b^{1/3}. \label{eqn:Tc}    
\end{align}
where $k_B$ is the Boltzmann's constant, and $\bar{\omega}_b$ is the geometric mean of the bosons' trapping frequencies. 

In contrast, for fermionic matter the Pauli exclusion principle prevents the simultaneous occupation of the same state by two fermions. As a consequence, at low temperature, every fermion settles into the lowest available energy state, successively filling them and forming a Fermi sea. The energy of the highest filled state is called the Fermi energy ($E_F$) and the corresponding temperature for this energy is the Fermi temperature $k_B T_F= E_F$. For a fermi gas containing \(N_f\) particles in a harmonic trap we have the following expression for $T_F$ \cite{pethick2008bose},
\begin{align}
    k_B T_F =  1.82 \hbar \bar{\omega}_f N_f^{1/3},
\end{align}
where \(\bar{\omega}_f\) is the geometric mean of the fermions' trapping frequencies.

We will refer to \(T/T_F\) and \(T/T_C\) as the reduced temperatures of the fermi and bose gases respectively. It is of relevance to note that for our experiment the masses and magnetic moments of \HeThreeStar{} and \HeFourStar{} give a fixed ratio of the trapping frequencies  \(\bar{\omega}_f = \sqrt{\frac{4}{3}} \bar{\omega}_b\) and hence,
\begin{align}
    T_F = 2.23 \left(\frac{N_f}{N_b}\right)^{1/3} T_C. \label{eqn:TC_TF}
\end{align}

\subsection{Momentum and Time-of-Flight distributions}
\label{sec:tof}

\begin{table*}[t]
\renewcommand{\arraystretch}{2.1}
    \centering
    \begin{tabularx}{\textwidth}{c|ccc}
        \toprule
    \toprule
         Statistics & In-trap Momentum Distribution & $n^{tof}(t)$ & $n^{tof}(x)$ \\
         \hline
         Bose-Einstein& $A_b \text{Li}_{\frac{3}{2}} \left[ \xi_b \exp \left(-\frac{p^2}{m_b^2 v_f^2} \right)\right]$ & $\pi m_{b}^3 A_b v_{b}^2 g \left(1+\frac{t_0^2}{t^2}\right) \text{Li}_{\frac{5}{2}}\left[ \xi_b g(t) \right]$& $\pi m_b^3 A_b \frac{v_b^2}{t_0}  \text{Li}_{\frac{5}{2}} \left[\xi_b f(x) \right]$  \\
         Fermi-Dirac&$-A_f \text{Li}_{\frac{3}{2}} \left[ -\xi_f \exp \left(-\frac{p^2}{m_f^2 v_b^2} \right)\right]$ &$-\pi m_{f}^3 A_f v_{f}^2 g \left(1+\frac{t_0^2}{t^2}\right) \text{Li}_{\frac{5}{2}} \left[ -\xi_f g(t) \right]$& $-\pi m_f^3 A_f \frac{v_f^2}{t_0}  \text{Li}_{\frac{5}{2}} \left[-\xi_f f(x) \right]$ \\
         Maxwell-Boltzmann& $A_t e^{-\frac{p^2}{m^2 v^2}}$ & $ \pi m_f^3 A_t  v^2 g \left(1+\frac{t_0^2}{t^2}\right) g(t)$ & $\pi A_t m_f^3 \frac{v^2}{t_0} f(x)$\\
         \hline
         Bose-Einstein Condensate & - & $\lambda A_{bec} t^2 \left[1- \left( \frac{(g(t^2-t_0^2)}{2R(t)}\right)^2 \right]$& $A_{bec} t_0^2 \left[1- \left( \frac{x}{\lambda R(t_0)}\right)^2 \right]$ \\
         \bottomrule
    \bottomrule
    \end{tabularx}
    \caption{Summary of the momentum and time-of-flight profiles for different particle statistics and a Bose-Einstein condensate. We have used the following abbreviations: $p$ is the magnitude of the three-dimensional momentum vector; \(v_{b,f} =  \Big(\frac{2k_BT}{m_{b,f}}\Big)^{1/2}\) is the most probable velocity of thermal bosons and fermions respectively; \(A_{b,f}=\frac{1}{(2\pi)^{3/2} \hbar^3} \left(\frac{k_B T}{m_{b,f} \bar{\omega}_{b,f}^2} \right)^{3/2}\), \(A_t = \frac{N_t}{(2\pi)^{3/2} (m k_B T)^{3/2}}\), and \(A_{bec}=\frac{\pi \mu_b^2}{m_b}\) are the normalization constants of the respective momentum distributions; $\xi_{f,b} = e^{\frac{-\mu_{f,b}}{k_B T}}$ is the fugacity of the respective distribution; \(\lambda=\frac{\pi \omega_x}{2\omega_\perp}\) is the aspect ratio of the time-of-flight profile of the BEC \cite{PhysRevLett.77.5315};
    \(R(t) = \sqrt{\frac{2\mu_b}{m_b}} t \); $g(t) =  \exp\left(-\frac{g(t^2-t_0^2)^2}{2v^2 t^2}\right) $; $f(x)=\exp\left(-\frac{x^2}{v^2t_0^2} \right)$. These distributions are all valid for \(v_{f,b} t_0\ll l_0 \). This is equivalent to assuming the average expansion of the cloud is negligible compared to the fall distance, which for our experiment is a very good approximation. We use the semi-classical approach and assume the behavior of an ideal gas for all distributions except the BEC. 
    For the BEC, where the mean-field energy dominates we use the scaling solution derived in Ref.~\cite{PhysRevLett.77.5315}.}
    \label{tab:tof}
\end{table*}

We analyze our bose-fermi mixture by measuring the time of flight profiles of each species on a micro-channel plate and delay line detector, which gives the position and time of each atom that strikes it (as described in Sec.~\ref{sec:detection}). In this section we will summarize the analytical forms of the time-of-flight and momentum profiles for both bosons and fermions. While the fermions' distribution can be completed described by a Fermi-Dirac statistics, the distribution of the bosons has two separate components, a BEC and non-condensed part which follows Bose-Einstein statistics. We will approximate this non-condensed component as thermal, and it will hence serve as our thermometer for the system.
 
The time-of-flight expansions of a BEC has been investigated by Castin et al.~\cite{PhysRevLett.77.5315}, and was found to be well approximated by a rescaling of the in trap density profile. If we have a cylindrical trap \(\omega_x \ll \omega_y,\, \omega_z\) and \(\omega_y=\omega_z=\omega_\perp\), we can solve for the long time, \(\omega_\perp t \gg 1\) where \(t\) is the expansion time, time-of-flight profile of the BEC for a harmonic potential to be,
\begin{align}
    n^{tof}_b(x,y,z,t) &= \begin{cases}\frac{\mu_b}{g_b} \left(1-\sum_{d=x,y,z} \frac{r_d^2}{R_d(t)^2}\right),\, \, \, r_d<R_d \\
    0, \, \, \, \text{otherwise}
    \end{cases}
\end{align}
where \(\mu_b\) is the boson chemical potential, \(g_b = \frac{4\pi \hbar^2 a_{44}}{m_b}\) is the interaction coefficient between bosons, with \(a_{44} = 7.512(5)\)~nm the \HeFourStar{}-\HeFourStar{} scattering length in the \(m_J=+1\) state \cite{PhysRevLett.96.023203}, \(R_{y,z}(t) = \sqrt{\frac{2\mu_b}{m_b}} t \) and \(R_{x}(t) = \frac{\omega_x}{\omega_\perp}\frac{\pi}{2} \sqrt{\frac{2\mu_b}{m_b}} t \). Note that the time-of-flight profile of a BEC does not ballistically map to its momentum profile \cite{PhysRevLett.76.6}, as is the case for the Fermi-Dirac and Maxwell-Boltzmann gases. Hence, generally the BEC profile alone does not encode any information on the temperature of the system. 

To determine the density distribution of the trapped cloud of fermions we take the semiclassical approximation (also known as the Thomas-Fermi approximation) \cite{PhysRevA.55.4346}. In this approximation each fermion is treated as having a definite position and momentum.
If we neglect interactions between the fermions due to the suppression of \(s\)-wave scattering of identical fermions, the ground state momentum density profile is given in this approximation by \cite{demarco2001quantum},
\begin{align}
    n_f(p) &= \frac{1}{(2\pi)^3} \int d^3 \vec{r}\, w(\vec{r},\vec{p})  \\
    &= -A_f \text{Li}_{3/2} \left[ -\xi \exp \left(-\frac{|\vec{p}|^2}{2m_f k_B T} \right)\right]
\end{align}
where \(\xi = e^{\frac{\mu_f}{k_B T}}\) is the fugacity, \(A_f=\frac{1}{(2\pi)^{3/2} \hbar^3} \left(\frac{k_B T}{m \bar{\omega}_f^2} \right)^{3/2}\), and \(\text{Li}_n\) is the poly-logarithmic function of order \(n\). 

At high momentum both the fermion and non-condensed boson distributions are well approximated by a classical Maxwell Boltzmann distribution
\begin{align}
    n_t(p) = A_t e^{-\frac{p^2}{2m k_B T}},
\end{align}
where \(A_t = \frac{N_t}{(2\pi)^{3/2} (m k_B T)^{3/2}}\), for a given number of thermal particles \(N_t\) at a temperature \(T\).

As we can neglect interactions in the fermionic and thermal components, the momentum distribution can be converted to the far-field time-of-flight profiles using the procedure described in Yavin et al. \cite{Yavin2002}. To represent the coordinates of our measurement procedure we will assume our detector is along the \(x-y\) plane, with the origin of the plane directly beneath the initial position of the cloud. Hence, the coordinates \((x,y,t)\) correspond to an atom arriving at the detector at time \(t\) at the spatial location \((x,y)\) on the detector. First we transform the momenta \((p_x,p_y,p_z)\) into time of flight coordinate \((x,y,t)\) via the equations for a ballistic motion under gravity 
\begin{align}
    (p_x,p_y,p_z) \mapsto m\times \left(\frac{x}{t},\frac{y}{t},\frac{g(t^2 - t_0^2)}{2t}\right),
\end{align}
where \(t_0=\sqrt{2l_0/g}\) is the fall time for a zero momentum particle, $l_0$ is the distance from the center of the trapped cloud to the detector and \(g\) is the acceleration due to gravity. We must also account for the change in differential volume between the two coordinate systems by multiplying the probability density by the relevant Jacobian \(J = m_{f,b}^3 \frac{gt^2 + 2l_0}{2t^4}\) \cite{Yavin2002}. Assuming the initial cloud can be treated as a point source, i.e.\ the spatial distribution can be ignored, the probability density distributions are related by
\begin{align}
    n^{tof}(x,y,t) = m_{f,b}^3 \frac{gt^2 + 2l_0}{2t^4} n\left(\frac{x}{t},\frac{y}{t},\frac{g(t^2 - t_0^2)}{2t}\right), 
\end{align}
where \(n(p_x,p_y,p_z)\) is the corresponding in trap momentum distribution.

To simplify the fitting process we integrate out two of the time-of-flight dimensions. Explicitly this can be written as
\begin{align}
    n^{tof}(t) &= \int^\infty_{-\infty} \int^\infty_{-\infty} dx \, dy \, n^{tof}_f(x,y,t) \\
    n^{tof}(x) &= \int^\infty_0 dt \, \int^\infty_{-\infty} dy \, n^{tof}_f(x,y,t).
\end{align}
The relevant momentum distributions, along with corresponding integrated time-of-flight profiles are displayed in Tab.~\ref{tab:tof}. Note that while the momentum profiles, and consequently the time-of-flight profiles, for the bosons and fermions look superficially similar their behavior is quite different, most relevantly in how rapidly their high momentum wings decay to a thermal distribution \cite{dist_note}.

\subsection{Laser trapping and cooling of \HeThreeStar{} and \HeFourStar{}}
\label{sec:levels}

\begin{figure}
\begin{center} 
    \includegraphics[width=\textwidth]{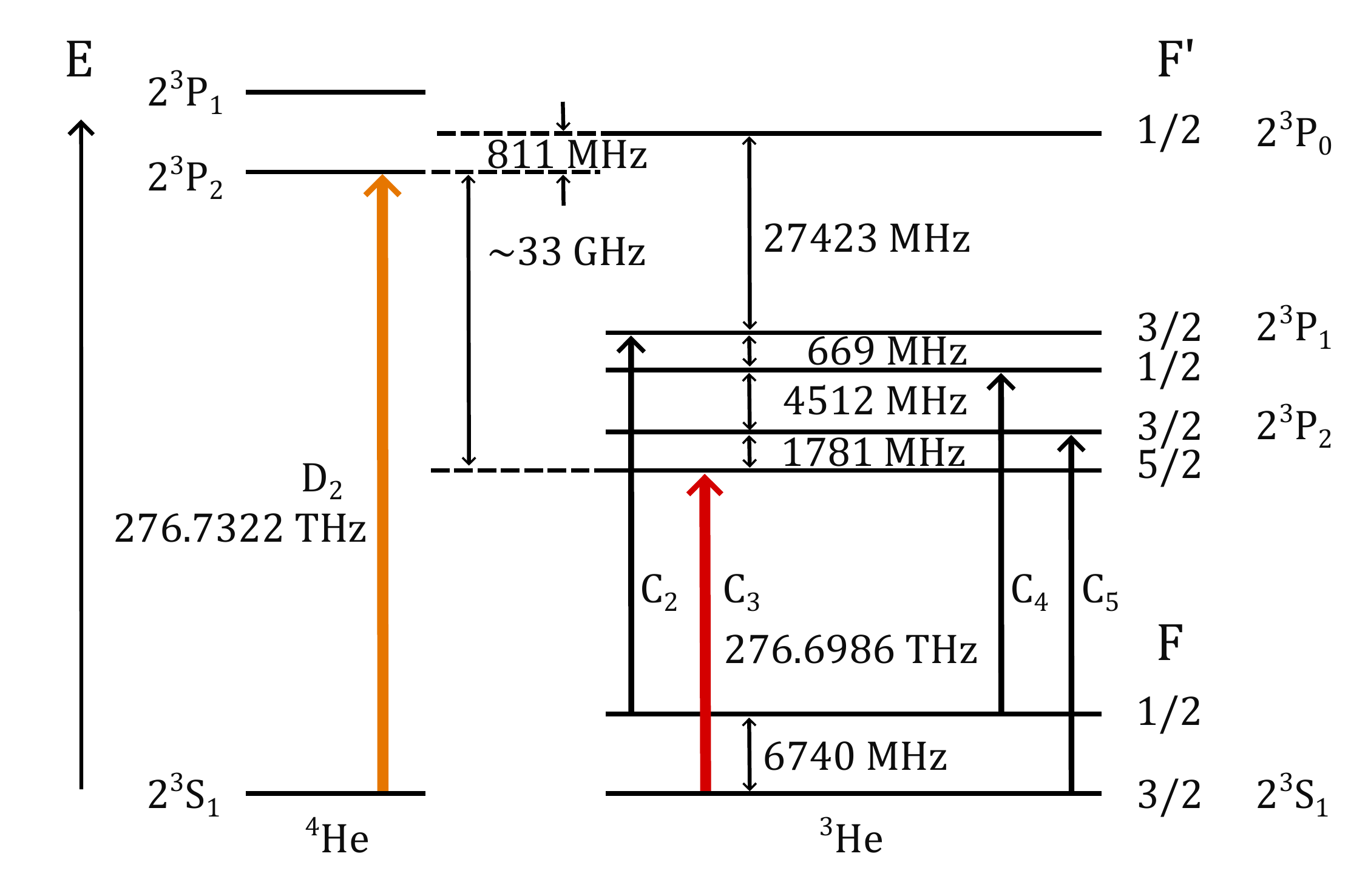}
  \end{center}
  \caption{Energy level diagrams of \HeFourStar{} and \HeThreeStar{} relative to the metastable state energy. The spacing of energy levels is only indicative. The 2\ch{^3S1} state of \HeFourStar{} has a lifetime of 7870(510) seconds~\cite{PhysRevLett.103.053002}. The $D_2$ and $C_3$ atomic transitions for laser cooling and trapping are indicated in orange and red respectively. The resonant frequencies of these two transitions are 276.7322~THz (1083.331~nm)~\cite{PhysRevLett.92.023001} and 276.6986~THz (1083.462~nm)~\cite{PhysRevA.32.2712}. Off-resonant excitation of the $C_5$ transition can be induced by the $C_3$ frequency, which is 1.781~GHz below the $C_5$ frequency~\cite{PhysRevA.32.2712}. However, the $C_2$ and $C_4$ act as repumpers to repopulate the $F$=3/2 metastable state \cite{Stas_phdthesis}. The hyperfine energy intervals in 2\ch{^3 P} of \HeThree{} are provided in \cite{PhysRevA.32.2712} and the interval in 2\ch{^3 S1} is given in Ref.~\cite{PhysRevA.1.571}.}
  \label{energy diagram}
\end{figure}

To achieve quantum degeneracy in our Bose-Fermi mixture of \hestar atoms we apply a series of laser cooling and trapping techniques to both \HeFourStar{} and \HeThreeStar{} in the $\MetastableState$ state simultaneously. The $\MetastableState$ state is a metastable state, with a lifetime of 7870(510) seconds for the \HeFourStar{} isotope~\cite{PhysRevLett.103.053002} and can therefore be treated as an effective ground state. The lifetime of the $\MetastableState$ state in \HeThreeStar{} is expected to be similar. From this state, the laser (Doppler) cooling transitions utilized in our experiment are the 
    $D_2\mathrm{:}~ \MetastableState \: (J=1) \rightarrow \CoolState \: (J=2)$ 
transition in \HeFourStar{} and the 
    $C_3\mathrm{:}~ \MetastableState \: (F=3/2) \rightarrow \CoolState \: (F=5/2)$ 
transition in \HeThreeStar{}, as highlighted in Fig.~\ref{energy diagram}. The Doppler cooling limit for both of these transitions is \(\sim\)39~{\textmu}K~\cite{jeltesphdthesis}. 

Efficient laser cooling is possible in both isotopes as these cooling transitions are \textit{closed}, with a high probably of returning the atom to the initial state after a cycle of photon absorption and decay.
In each case a photon absorption initially populates the $\CoolState$ state which predominately decays to the metastable state as a decay to the ground state is forbidden~\cite{Hodgman2009a}. 
In the hyperfine structure of \HeThreeStar{} ($I=1/2$), the excited $F=5/2$ state can only decay to the $F=3/2$ metastable state following the selection rule $\Delta F=\pm1$. 
In the presence of a magnetic field the (initial) total angular momentum projection may also be maintained after a laser cooling absorption-emission cycle. Here $\sigma^+$-polarized cooling light is used to drive the $m_J=1 \rightarrow m_J=2$ transition in \HeFourStar{} ($m_F=3/2 \rightarrow m_F=5/2$ transition in \HeThreeStar{}) which can only decay to the original state given the $\Delta M_{F,J}=\pm1$ selection rule.

While the $C_3$ cooling transition in \HeThreeStar{} itself is closed, the small energy differences between the excited hyperfine states~\cite{PhysRevA.32.2712} comparable to their natural linewidths ($\Gamma\approx$1.6~MHz~\cite{RevModPhys.84.175}), means that laser cooling at this frequency will drive other excitations. 
Specifically there will be a weak off resonant excitation from the metastable state to the $\CoolState$ ($F'=3/2$) state ($C_5$ transition), from which the atom decays to the $\MetastableState$ $F=1/2$ state, resulting in a depletion of the $F=3/2$ metastable state.
However, this depletion is countered by the stronger off resonant driving of transitions which drive population out of the $F=1/2$ state.
In particular as the detunings of the $C_3$ laser with respect to the $C_2$ and $C_4$ transitions are -249$\Gamma$ and 169$\Gamma$ respectively (c.f. $\sim 1100 \Gamma$ for $C_5$), the $F$=3/2 metastable state is quickly repopulated by decay from the $\CoolState \: (F'=1/2$) or $2^{3\!}P_1 \: (F'=3/2$) states \cite{Stas_phdthesis}. 
Therefore, this state leakage is strongly suppressed and an extra laser frequency is not required to excite atoms out of the `dark' ground state for any stage of our experiment.

The detuning between the $D_2$ and $C_3$ transitions used to cool each isotope is sufficiently large (33.574~GHz~\cite{PhysRevA.32.2712}) that there is negligible off-resonant excitation of one isotope from the others cooling laser.

\subsection{Evaporative and Sympathetic Cooling}
\label{sec:evap}

For our system laser cooling alone is not enough to reach quantum degenerate temperatures. We have previously successfully achieved degeneracy in \HeFourStar{}, reaching BEC, by applying radio-frequency (RF) induced forced evaporation (on the \(2^3S_1\) \(m_J=+1\rightarrow m_J=0\) transition) in a magnetic trap \cite{Dall2007}. However, this procedure cannot be directly applied to fermionic \HeThreeStar{}. Only the \(s\)-wave component of scattering interactions has a nonzero contribution to the scattering amplitude in the low energy limit. Furthermore, due to the Pauli exclusion principle the \(s\)-wave scattering length of fermions is zero, and thus a fermi gas will be unable to rethermalize with itself via elastic collisions \cite{PhysRevA.61.013406}. However, as the fermions are in thermal contact with the bosons and the interspecies scattering length is relatively large (\(a_{34} = 29(4) \)~nm \cite{PhysRevA.104.033317}), for sufficiently slow evaporation the fermions will thus remain in thermal equilibrium with bosons as they are cooled. Given that the expected interspecies scattering length is significantly higher than the \HeFourStar{} intraspecies scattering length (\(a_{44}=7.512(5)\)~nm \cite{PhysRevLett.96.023203}) we were able to use the same evaporation cycle that is used for a pure sample of \HeFourStar{} (approximately \(16\)~s in length). The change in temperature of the mixture due to the forced evaporation process depends on the efficiency of the evaporation ramp (parameterized by \(\eta\)), the sum of the heat capacities, and the rate of change in energy of the Bose gas with respect to particle number \cite{PhysRevA.69.043611}. As the behavior of the thermodynamic properties of the gases changes between the classical and degenerate regimes, especially for a bose gas, the exact behavior is fairly complex. However, if the heat capacity of the fermi gas is negligible compared to the heat capacity of the bose gas for all temperatures greater than \(T_C\), then the final reduced temperature of the fermi gas \(T/T_F\) is only dependent on the ratio \(N_b/N_f\), where \(N_b\) and \(N_f\) are the number of bosons and fermions in the mixture after evaporation \cite{evap_note}. This implies that having a larger initial number of fermions is equivalent to evaporating closer to the trap bottom (the detuning for zero momentum), in terms of the reduced Fermi temperature (\(T/T_F\)) reached. Furthermore, as \(T_F\) and \(T_C\) are related by Eqn.~\ref{eqn:TC_TF} and the mixture is in thermal equilibrium, the reduced temperature of the bose gas will also only depend on the ratio \(N_b/N_f\). This also implies that for a given evaporation efficiency and initial phase space density there is an optimal Fermi reduced temperature that can be reached (i.e.\ when all the coolant, \HeFourStar{} gas, has been used up).

We expect a negligible amount \HeThreeStar{} to be removed during the evaporation process, even in the thermal regime where the fermionic scattering lengths are non-zero. The energy due to the external magnetic field for a given substate is \(E_{m_{F/J}}=g m_{F/J} \mu_B B\), where \(\mu_B\) is the Bohr magneton, \(B\) is the field strength, and \(g\) is the gyromagnetic ratio. If we consider the relevant transition for the magnetically trapped substates of each species (\(m_J=+1\rightarrow m_J=0\) for \HeFourStar{} and \(m_F=+3/2\rightarrow m_F=+1/2\) for \HeThreeStar{}) and their gyromagnetic ratios (\(g=2\) and \(g=4/3\) respectively) we can see that the resonant frequency for a given magnetic field is \(3/2\) times larger for \HeFourStar{} than it is for \HeThreeStar{}. For our trap minima of \(B=0.3\)~G this corresponds to a detuning of \(\sim 0.3\)~MHz. Thus, if \HeThreeStar{} is in thermal equilibrium with \HeFourStar{} then \HeFourStar{} will be preferentially evaporated, assuming we start our forced evaporation sweep above the \HeFourStar{} trap bottom. This effect is further exacerbated by the \HeThreeStar{} atoms experiencing a \(\sqrt{\frac{4}{3}}\) times higher trapping frequency due to their smaller mass and thus will sit closer to the trap center.

We characterize the evaporation cycle by the detuning, or height, of the final frequency applied from the frequency of the transition for an atom at zero temperature. This is equivalent to the detuning for an atom sitting at the minimum of the trapping potential, thus we refer to this parameter as the evaporation height above trap bottom. Note that once all \HeFourStar{} atoms have been evaporated, if we continue to lower the evaporator height then \HeThreeStar{} atoms will be removed.

\section{Apparatus}

\begin{figure}
\begin{center}  
    \includegraphics[width=\textwidth]{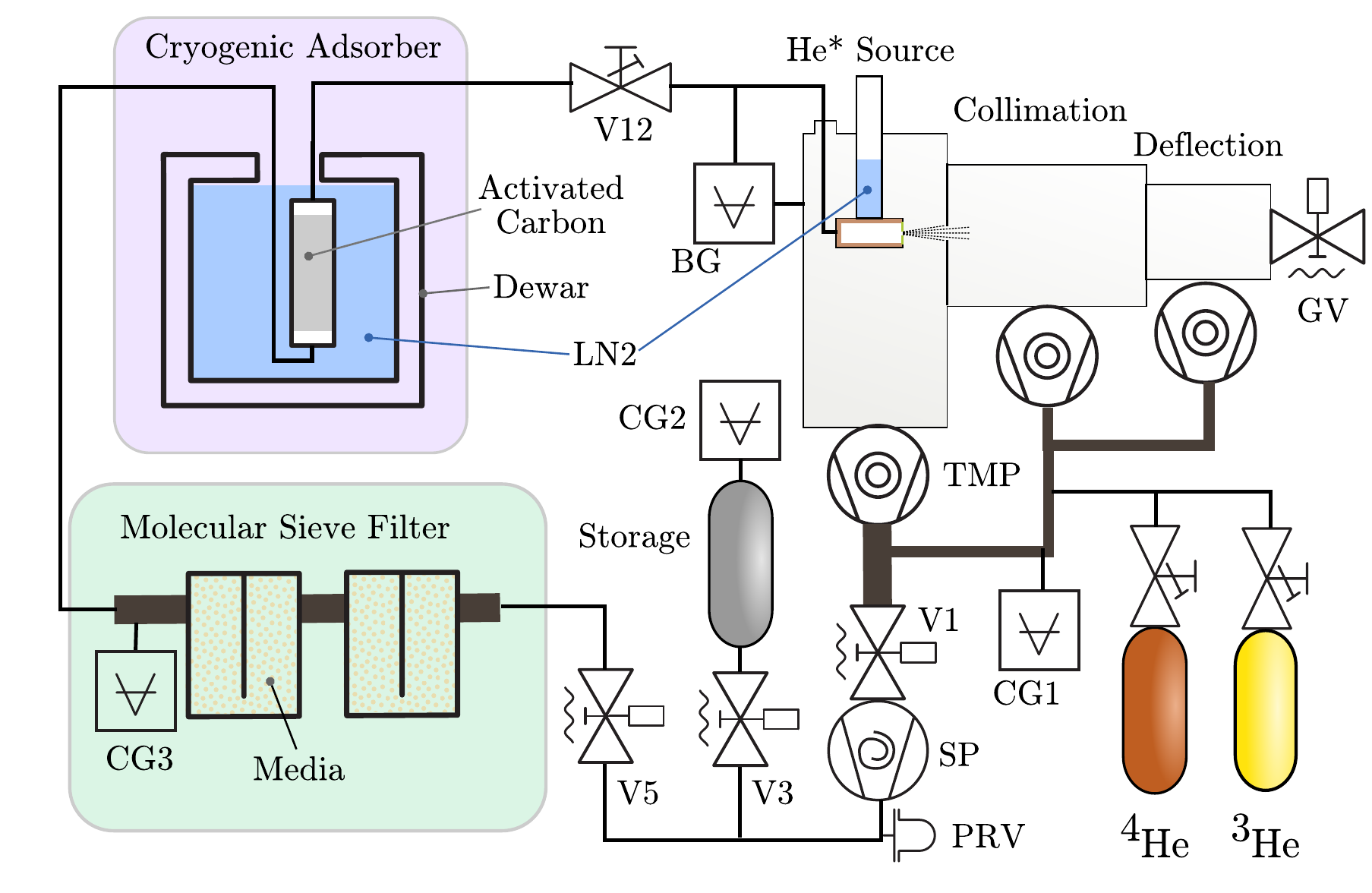}
  \end{center}
  \caption{
  Simplified schematic of the recycling system.
  Abbreviations:
  TMP - Turbomolecular Pump,
  SP - Scroll Pump,
  PRV - Pressure Release Valve,
  BG - Baratron Gauge,
  LN2 - Liquid Nitrogen,
  GV - Gate Valve,
  V - Valve, and
  CV - Convectron.
  }
  \label{fig:recirc_simple}
\end{figure}

In this section we will cover the technical aspects of our experimental apparatus, focusing on the modification required to incorporate \HeThreeStar{} into the cooling process, specifically the laser and vacuum system.

\subsection{Source of Metastable Atoms}

Before using laser cooling and trapping techniques, we need to first produce metastable helium. However, optically producing He$^*$ by exciting helium from the ground state to the 2\ch{^3S1} is impractical, as it would require far ultraviolet light, which is not conveniently accessible, and furthermore the scattering cross-section of the transition is extremely small~\cite{metcalf1999laser}. Therefore, we use high energy electron collisions in a liquid nitrogen cooled DC discharge plasma to excite both species to the 2\ch{^3S1} state. This plasma is focused by an insulating nozzle through with the gas flows before undergoing a supersonic expansion~\cite{swansson2004high}. Atoms with a low transverse velocity are then selected to enter the next stage with a simple mechanical aperture.

\subsection{Atom-optics Elements}

The atom beam comprising both \HeFourStar{} and \HeThreeStar{} is initially collimated by cooling in the transverse direction with a 2-D optical molasses and then slowed via a Zeeman slower, in which a red-detuned laser light combined with a magnetic field acts to continuously slow the beam. The slowed \hestar{} atoms are then captured in a magneto-optic trap (MOT) variant which produces a focused cold beam of \hestar, which is used to load a second MOT~\cite{LVIS}. After transferring the atoms from the MOT to a magnetic trap, 1-D Doppler cooling is employed to further cool the atoms~\cite{Dall2007}.

\subsection{Magnetic Trap}
We perform our 1D Doppler cooling and RF evaporation of the two species in a biplanar quadrupole Ioffe magnetic trap \cite{Dall2007}, which can be treated as a harmonic potential with cylindrical symmetry. For the data presented in this work we have used trapping frequencies for \HeFourStar{} of \(\{\omega_x, \omega_y, \omega_z\}/2\pi \sim  \{60(2),600.5(1),605.7(7)\}\)~Hz, with gravity aligned along the \(z\)-axis. The trapping frequencies were measured using a pulsed atom laser technique described in Ref.~\cite{Henson:22}.

\subsection{Detection}
\label{sec:detection}
We measure the far-field time of flight profiles for both species using a micro-channel plate (MCP) and delay line detector (DLD) \cite{Manning:10} system, which due to the high internal energy of both species metastable state allows us to detect the positions of single atom impacts after they are released from the trap. The MCP is located \(858.7\)~mm below the trap, which corresponds to a fall time of \(t_0 = 0.4187\)~s, for a particle with zero initial momentum.

As both clouds momenta distributions are initially centered on zero they will overlap on the detector if they are only influenced by gravity. To separate the clouds we apply a linear magnetic field gradient to both species, produced by a set of coils surrounding our BEC chamber. While the two isotopes have the same magnetic moment, and hence experiences the same force from the external field, they will be accelerated by different amounts due to their differing masses. This pulse alters our time of flight profiles and must be accounted for in order to obtain accurate measurements of the mixtures' properties, see Sec.~\ref{sec:tof}.
 
To avoid detector saturation \cite{Manning:10} we employ a pulsed atom laser technique for \HeFourStar{} \cite{Henson:22}, which out couples only small portions (\(\sim 2 \%\)) of the cloud, with each pulse being well below the saturation limit and thus giving an accurate measure of the \HeFourStar{} atom number. In order to determine the atom number of the \HeThreeStar{} cloud we employ absorption imaging. 
The linear polarization of our probe beam will pump the imaged atoms into a steady state after some time (approximately \(50\)~\(\mu\)s for our experiment). Using the relevant Clebsch-Gordan coefficients we can solve for the steady state distribution and find \(25/14\) increase in our saturation intensity for \HeThreeStar{}. Hence, we can find the number of \HeThreeStar{} atoms in our mixture via the relation \(N_f = \frac{1}{\sigma_a} \iint dx dy \, D(x,y)\), where the optical density \(D\) is defined as \(I_{\text{out}} = I_{\text{in}} e^{-D}\) and the absorption cross-section \(\sigma_a\) is the absorption cross-section \cite{doi:10.1119/1.12937}.

\subsection{Recirculation system}

\begin{figure}
\centering
\includegraphics[width=\textwidth]{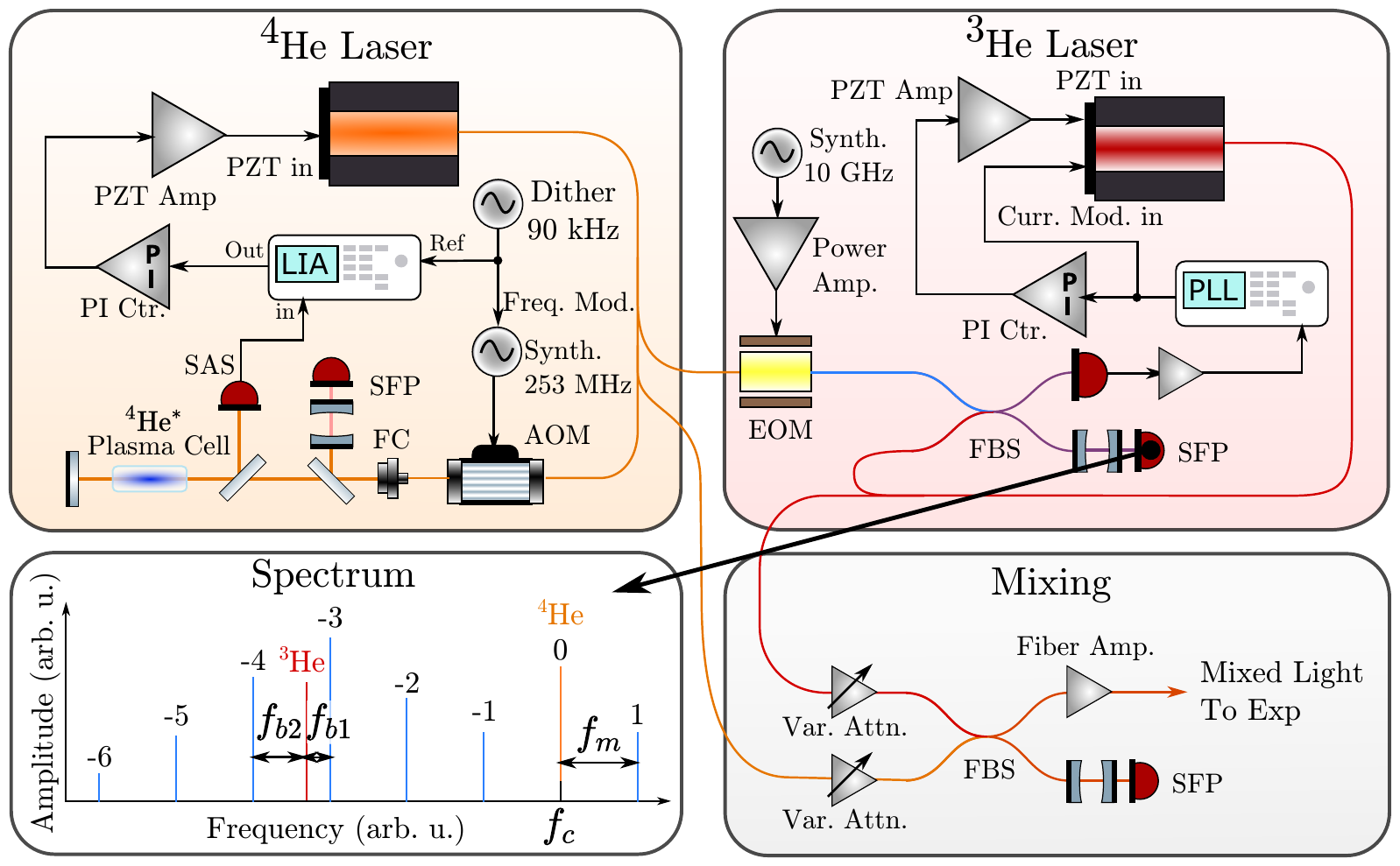}
\caption{
Laser system. Sidebands of the \HeFourStar{} laser are generated by an EOM, driven by MW frequency synthesizer. The \HeFourStar{} light, together with its sidebands, is mixed with the \HeThreeStar{} laser on a photodiode, synthesizing a series of beat signals. One of the beat notes is chosen as the input to the PLL, whose output controls the current modulation of the laser and is also sent to a PI controller. After being integrated by the PI controller, the output is fed back to the PZT of the \HeThreeStar{} laser to actively lock its frequency to the expected value. Abbreviations: AOM - Acousto-Optic Modulator, EOM - Electro-Optic Modulator, Var. Attn. - Electronic Variable Optical Attenuator, PZT - Piezoelectric Actuator, Curr. Mod - Current Modulation, Freq. Mod. - Frequency Modulation, Synth. 10~GHz - Windfreak MW Synthesizer, PI Ctr. - Proportional Integral Controller, SAS - Saturation Absorption Spectroscopy, FBS - Fiber Beamsplitter, SFP - Scanning Fabry-Perot Interferometer.
In the spectrum, with a phase modulation of 4.4~rad, the 3rd sideband amplitude is 0.22. The parameters $f_{b1}$ and $f_{b2}$ are the lowest frequency beatnotes produced when the $^{3}$He laser is mixed with the frequency comb generated with the EOM.
}
\label{ch:tune_out.fig:he3_laser_system}
\end{figure}

\begin{figure}
         \centering
         \includegraphics[width=\textwidth]{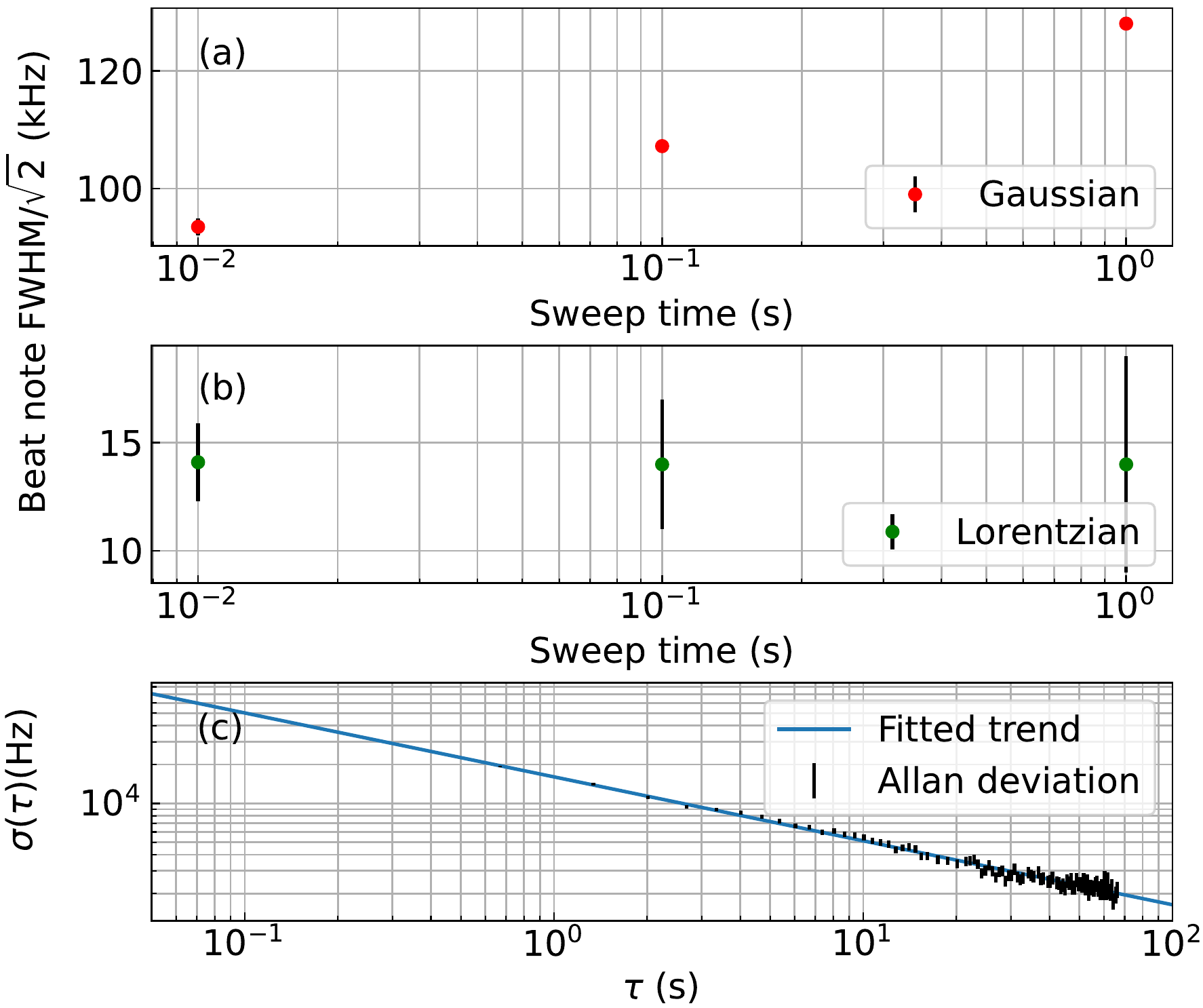}
         \hfill
         \caption{(a) and (b): FWHM/$\sqrt{2}$ of the Gaussian and Lorentzian of the locked beat note for different sweep times respectively. As the sweep time increases, the beat note linewidth given by the Gaussian shows a increasing trend while the linewidth estimated by the Lorentzian remains to be stable. The beat note is estimated to have a linewidth of $\sim$100~kHz. (c): Allan deviation $\sigma(\tau)$ of the beat note frequency. It decreases over time as the noise of the signal measured at different times cancels out. The fitted relationship between $\sigma$ and integration time $\tau$ is $\sigma(\tau)=16037\sqrt{\mathrm{~Hz}}\: \tau^{-1/2}$. 
         }
         \label{linewidth}
\end{figure}

The plasma discharge source used to create \hestar{} is relatively inefficient with only a small fraction ($10^{-8}$) of the helium atoms input into the source being converted into \hestar{} and making it into the first stage of the system (transverse cooling). The remaining gas is pumped away with a turbo molecular pump and previously had been discharged to atmosphere via the exhaust of a backing pump.

While this approach is reasonable with \HeFour{}, which is relatively inexpensive, the cost of \HeThree{} means that this approach would quickly become prohibitive given the relatively high flow rate (0.7(1) NTP L/h) required by the source. We have therefore implemented a system which can recycle this helium gas while maintaining the high purity required for efficient metastable producible.

The plasma source of \hestar{} is quite sensitive to contaminants owing to both gas interactions and mechanical effects.  
First in a gas of helium and metastable helium the presence of impurities acts to destroy the metastable component through a process known as  penning ionization. 
This process is highly favored for almost all impurity atoms (except for neon) given the large helium metastable state energy compared to the impurities' ionization energy.
Furthermore, contaminants can mechanically degrade the source; organics such as oil are decomposed by the source into carbon soot, while impurities such as oxygen accelerate the erosion of the copper and tungsten materials of the source.
In the previous system contaminants were avoided by using high purity helium gas, however in the recycling system oil from the vacuum pumps and the inevitable leakage of air into the system will accumulate and must be dealt with. 

To recycle the gas requires compressing the gas that is captured by the three relevant turbo vacuum pumps (see Fig.~\ref{fig:recirc_simple}), to the pressure needed to run the discharge source (between 10 and 0.5 kPa) and then feeding it back into the source.
To reduce the source of oil as much as possible we replaced the rotary-vane backing pump for the source turbo, which uses oil as a sealing and lubricating agent, with a `dry'(oil-free) scroll pump. 
This does not eliminate oil from the system entirely as all three of the turbomolecular pumps use oil lubrication of the rear bearing. 
To manage this contamination two filter stages are used. First, on the backing line of the turbomolecular pumps we use micromaze traps comprising a number of highly porous ceramic plates that trap oil vapor. 
This serves to reduce the contamination of the scroll pump which compresses the gas to the source pressure. 
After compression the gas then passes through two molecular sieve filters filled with SSZ-13 zeolite media. 
This second stage further reduces oil and presents a large surface area which also serves to adsorb water vapor. 
The molecular sieves can be partially regenerated by removing the circulating gas and heating them to approximately 120{\textdegree}~C. 

As the recycling system is closed, any small air leaks will compound over time.
To remove this air the final filtration stage uses a liquid nitrogen cooled activated carbon adsorber. 
At low temperatures gas molecules make a weak van der Waals bond to the carbon surface through a process known as adsorption which is maximized by using activated carbon with a large surface area.
Operation of the adsorber at liquid nitrogen temperatures is able to remove all atmospheric gases (oxygen, nitrogen, water) while preventing adsorption of helium. 
To regenerate the adsorber the liquid nitrogen is removed from the surrounding dewar and the unit heated to approx 120~{\textdegree}C. During regeneration of the adsorber and the molecular sieves the helium gas is stored in a separate storage tank.

A key parameter for engineering this system is the amount of \HeThree{} needed to fill the system. This is set by the volume (tubing and filters) which is pressurized to the pressure required to start the source plasma. This pressure is decreased (compared to our previous design) to a few kPa by using a flyback based high voltage ignition system from a residential barbecue.  The total volume at the source pressure is kept to 5~L, while the total gas need to operate the system is then 0.1 L STP, which in a 50:50 ratio of \HeThree{} to \HeFour{} represents a few hundred AUD of \ch{^3 He}. 

The recycling system only captures gas before the Zeeman slower. However, the loss of helium which goes beyond this point from both the helium background gas pressure ($1\cdot10^{-2}$~L STP/year) and metastable flux ($2\cdot10^{-4}$~L STP/year) are acceptably small.

As a test of operation, \HeFour{} has been successfully recirculated for months at a time without problem. A residual gas analyzer fixed to the collimator stage was used to monitor the gas composition of the source, with no detectable impurities above the background levels in the chamber.

\subsection{Laser lock}

A laser with a stable frequency is important for cooling and trapping an atomic gas. In our setup an external-cavity laser (ECL) system is used for \HeFourStar{}~\cite{ECL}, featuring a diffraction grating that selects the coarse laser frequency.
The grating is installed on a piezoelectric chip (PZT) that changes the cavity length and finely modulates the frequency. This ECL is locked to the $D_2$ transition in \HeFourStar{} using saturated absorption spectroscopy (SAS)~\cite{ECL}. For \hestar{}, the need to have an excited state population, means that SAS requires an RF discharge driven plasma inside the vapor cell as an atomic reference. Due to problems caused by the plasma etching, the cell needs to be refilled regularly. However, because of the high cost of \HeThree{} gas we have utilized an alternative setup, based on a frequency comb and a phase-locked loop, which locks the \HeThreeStar{} laser relative to the \HeFourStar{} laser by stabilizing the beat frequency between the two lasers.

As discussed in Sec.~\ref{sec:levels}, the \HeThreeStar{} laser is locked to the $C_3$ transition ($2^{3\!}S_1 (F=3/2 )\rightarrow 2^{3\!}P_{2} (F'=5/2)$), which is 33.574~GHz red detuned from the \HeFourStar{} $D_2$ transition ($2^{3\!}S_1 \rightarrow 2^{3\!}P_{2} $). This detuning is too large for conventional electronics to easily process. To circumvent this we used a high frequency optical phase modulator(ixblue NIR-MPX-LN-10) to create a small span frequency comb, with the \HeFourStar{} laser as the carrier frequency ($f_c$) and tooth spacing $f_m$ set by the modulation frequency~\cite{Peng:14,Harada:16}. Specifically the \(N\)-th order tooth (or sideband) has the frequency $f_c\pm Nf_m$. For our experiment $f_m\sim$10~GHz and is set by a microwave synthesizer (Windfreak SynthHD v2). 

The \HeThreeStar{} laser is then mixed with light from the frequency comb on a photodiode, generating a series of beat frequencies equal to the absolute value of the offset between the \HeThreeStar{} laser and the sidebands.
The lowest of these beatnotes ($f_{b1}=|f_{ ^3 He}-(f_c+ Nf_m)|$) is for the third-order lower sideband $N=-3$ in our system, and is at most 5~GHz.
The beatnote can be reduced to arbitrarily small values via tuning $f_m$, for our experiment $f_{b1}\sim$1~GHz. 
The value of $f_{b1}$ is actively stabilized by a feedback loop that compares it with a reference frequency using a Fractional-N phase frequency detector (Linear Technology LTC6947). The LTC6947 board outputs a signal proportional to the phase and frequency difference between $f_{b1}$ and the target frequency. The error signal is then used to stabilize $f_{b1}$ through both a fast proportional-only path to the \HeThreeStar{} laser diode current and a slower proportional-integral~(PI) path which controls the PZT of the \HeThreeStar{} laser. 
As $f_c+ Nf_m$ is fixed stabilizing $f_{b1}$ is equivalent to stabilizing $f_{ ^3 He}$.
In this system we find the frequency difference between the \HeThreeStar{} $C_3$ transition and $f_{ ^3 He}$ ($\Delta_{C3}$) using the previously measured frequency difference of $D_2$ and $C_3$, $\Delta_{C3}=f_{ ^3 He}-f_{ ^4 He}+33\:574$~MHz \cite{PhysRevA.32.2712}. 
This procedure is illustrated in Fig.~\ref{ch:tune_out.fig:he3_laser_system}, which also shows an indicative spectrum of the frequency comb.

The linewidth of the beat note is one of the primary indicators of the performance of the laser lock. To evaluate the linewidth, a Gaussian is fitted to the central region that spans $\approx$1~MHz of its spectrum and a Lorentzian is fitted to the tails~\cite{ThompsonandScholten,linewidth}.  
Figure \ref{linewidth} shows the linewidths measured for a series of sweep times of the spectral analyzer. The linewidth is deduced to be on the order of 100~kHz, which is larger than the linewidth of the \HeFourStar{} laser ($\sim$33~kHz~\cite{ECL}), as is expected, but much smaller than the natural linewidth of the relevant transitions ($\Gamma\approx1.6$~MHz).
The Allan deviation $\sigma(\tau)$ (shown in Fig.~\ref{linewidth}) indicates that the laser could readily be used in spectroscopic measurements, with a measurement uncertainty of $\sim$5~kHz only requiring 10 seconds of integration time. 
This system could in principle be used to stabilize even larger frequency offsets by increasing the modulation drive power and using higher modulation orders. It could also lock multiple lasers at the same time as the maximum beatnote frequency (5~GHz) can be readily captured on a photodiode.

Incorporating \HeThreeStar{} light into the existing \HeFourStar{} BEC experimental apparatus \cite{Dall2007} required overlapping the \HeThreeStar{} laser with the \HeFourStar{} laser for every atom-optic element of the apparatus.
Previously, cooling light for \HeFourStar{} was produced by amplifying light from the master laser with a fiber amplifier (Nufern, NUA-1064-PB-0005-A2). To cool both \HeThreeStar{} and \HeFourStar{} we seed this fiber amplifier with a mixture of $C_3$ and $D_2$ cooling light.
The output of the amplifier then contains both frequencies of light, which is overlapped in the atom-optic elements previously arranged for the \HeFourStar{} laser. 
This avoids duplicating many of the optical component of \HeThreeStar{}. 
The ratio of the frequency components at the output is then dynamically controlled using electronically variable attenuators on the inputs. 
We have found that the spatial overlap for the frequency components along the optical path is robust even with the incision of AOM's. 
In this approach the independent control the detuning of \HeThree{} and \HeFour{}  cooling light for each atom-optic stage via the AOM is compromised for the simplicity and convenience of overlapping the two lasers using the fiber amplifier. 
As we only need to sympathetically cool a limited amount of \HeThree{}~\cite{PhysRevA.69.043611}, this compromise only leads to a minimal decrease from the optimal conditions that would be possible with independent optics for each laser.

\section{Results}

In this section we will discuss the level of degeneracy achieved in our Bose-Fermi mixture, along with how our main two control parameters, detuning of the \HeThreeStar{} laser and final stage RF evaporator height, affect the most relevant experimental parameters, reduced temperature and atom number of the respective gases.

\subsection{Time-of-flight profiles}

\begin{figure}[t]
    \centering
    \includegraphics[width=\textwidth]{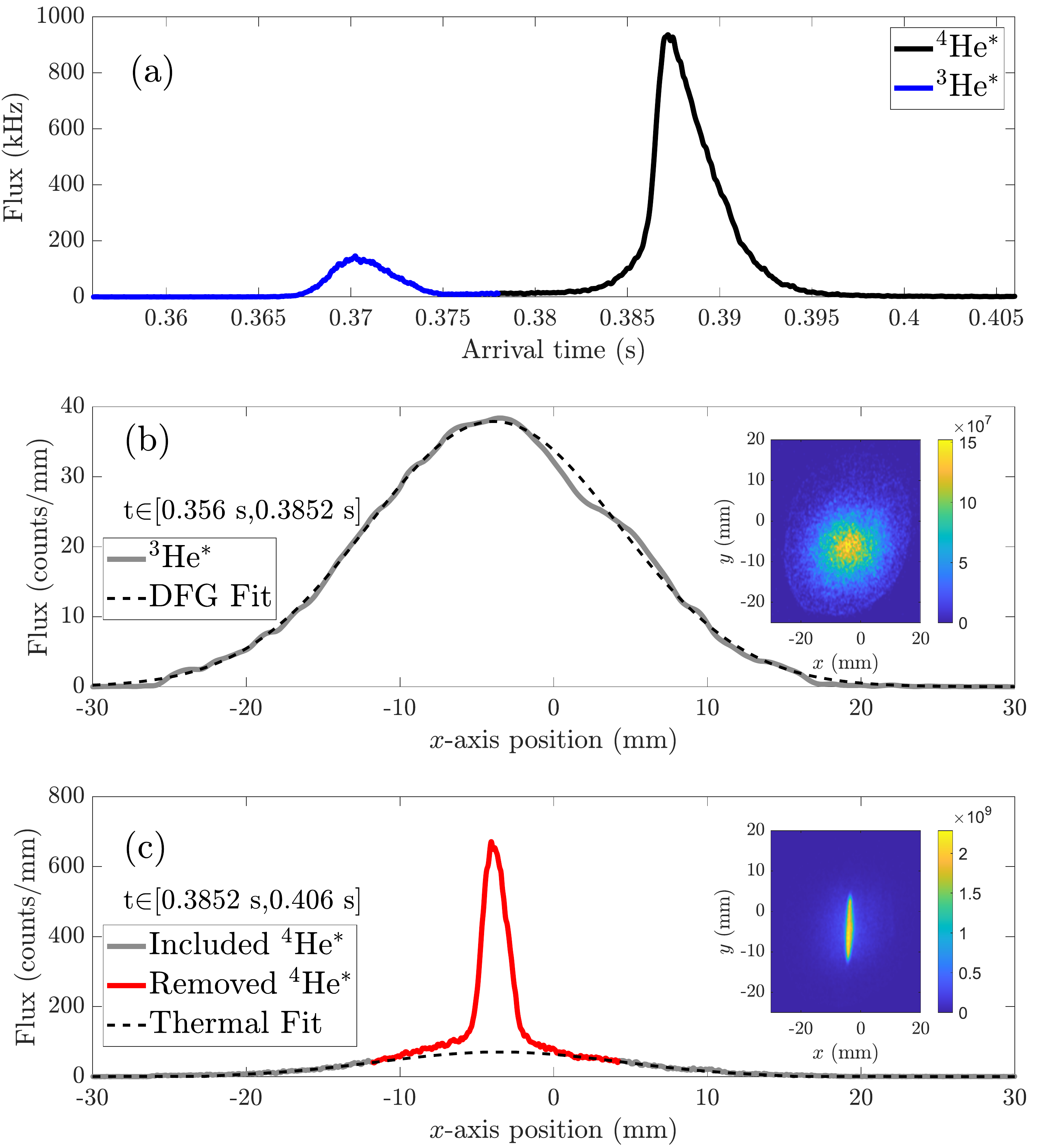}
    \caption{Time-of-flight profiles for \HeThreeStar{} and \HeFourStar{} for a final evaporation height of \(15\)~kHz and detuning\(\Delta_{C3} = 1.0\)~MHz. (a) We show the flux as a function of arrival time after trap switch off. The two species are separated in time by using a magnetic field gradient. The \HeThreeStar{} cloud lands on the detector first and is highlighted in blue, while the center of the \HeFourStar{} cloud arrives 10~ms later. Some saturation of the BEC is present in the \HeFourStar{} profile, however this is not an issue as we do not derive any information from this section of the profile. (b) Average \(x\)-axis flux of the \HeThreeStar{} cloud fitted with a Fermi-Dirac distribution with \(T = 174(10)\)~nK, and \(\xi=3.8(2)\) (black dashed line). Inset shows the density distribution over the \(x\)-\(y\) plane, demonstrating the isotropy of the cloud. (c) Average \(x\)-axis flux of the \HeFourStar{} cloud. We fit a thermal distribution (black dashed line) with \(T=215(18)\)~nK to the wings of the distribution, coloured in gray, removing the central area which is either overlapped with the condensate or affected by its mean-field, coloured in red. We fit along the \(x\)-axis as this is has the smallest BEC momentum width. This corresponds to a reduced temperature of \(T/T_F=0.40(4)\) for the fermions and \(T/T_C=0.29(5)\) for bosons. Inset shows the density distribution over the \(x\)-\(y\) plane, demonstrating the anisotropy of the BEC.
    }
    \label{fig:tof_profile}
\end{figure}

We find the temperature of the final mixture from the high momentum tails of the time-of-flight profile of the \HeFourStar{} atoms (see Fig.~\ref{fig:tof_profile}). This is due to a number of factors, however, it is primarily because it is difficult to accurately fit the temperature and fugacity of Fermi-Dirac distribution independently as there is strong cross correlation between their values, significantly stronger than in the Bose-Einstein distribution (see Ref.~\cite{dist_note} for further details). However, we still fit the Fermi-Dirac distribution as a verification of the order of magnitude of the temperature measurement of the \HeFourStar{} cloud.  


This approach of thermometry using only the \HeFourStar{} cloud assumes that both gases are in thermal equilibrium at the time of release from the trap. Due to the high scattering length between the species \(a_{34}=29(4)\)~nm \cite{PhysRevA.104.033317} we expect them to thermalize relatively quickly. However, due to the difficulty of extracting the temperature from the Fermi distribution accurately we cannot confirm this directly. In the data presented here we hold the clouds for 400 ms after the end of evaporation.

\begin{figure}[t]
    \centering
    \includegraphics[width=\textwidth]{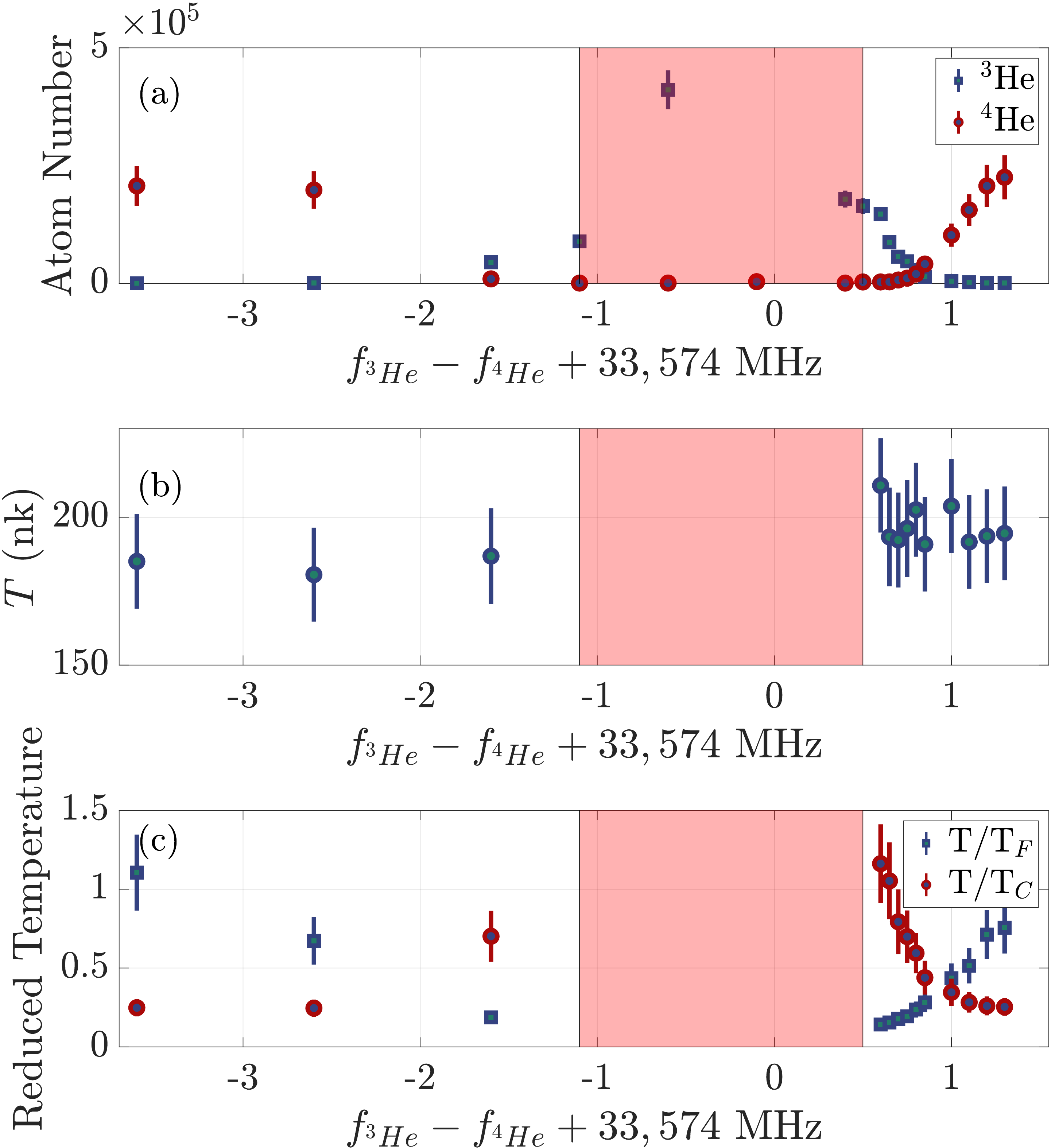}
    \caption{Atom numbers (a) temperature (b) and reduced temperature (c) of both species of the Bose-Fermi mixture against the detuning of the \HeThreeStar{} laser from the cooling transition ($\Delta_{C3}$) for a fixed evaporator height of \(15\)~kHz. This allows us to effectively load various amounts of \HeThreeStar{} atoms into the trap, while leaving the initial number and temperature of the \HeFourStar{} atoms unchanged. The region with no \HeFourStar{} atoms is shaded red. We do not present temperature measurements in this region as our temperature measure from the Fermi cloud is unreliable.}
    \label{fig:detuning_scan}
\end{figure}

\subsection{Dependence on laser detuning and evaporator height}

\begin{figure}[t]
    \centering
    \includegraphics[width=\textwidth]{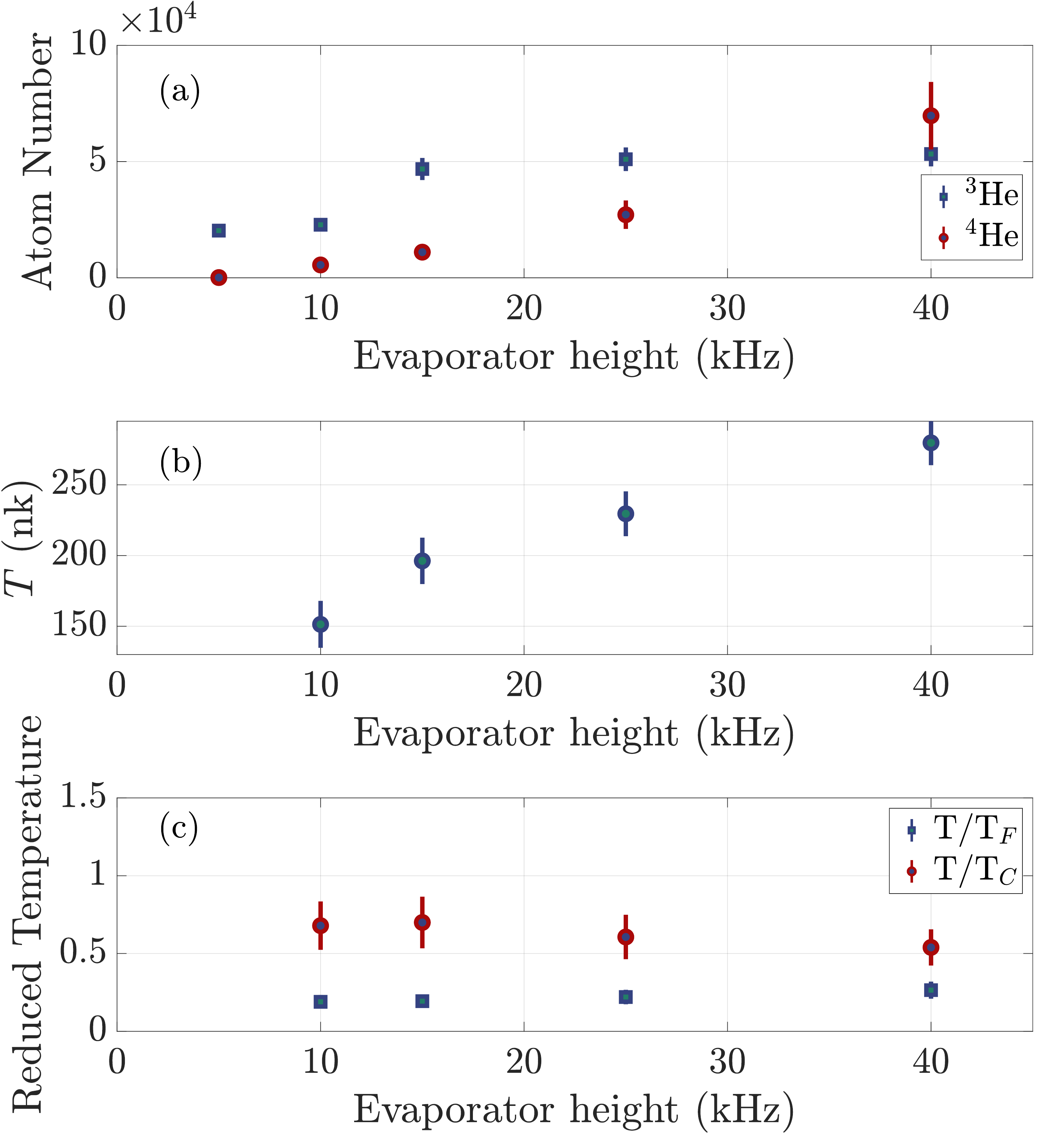}
    \caption{Atom numbers (a), temperatures (b), and reduced temperature (c) versus the final height of the forced evaporation above the trap bottom, for a fixed \HeThreeStar{} detuning of \(\Delta_{C3}=0.75\)~MHz. Note that the \HeThreeStar{} number remains unchanged until almost all the \HeFourStar{} have been evaporated, as expected. Furthermore, we see that while the temperature of the mixture decreases with evaporator height, the reduced temperatures remain roughly constant, with only a slight reduction in \(T/T_F\).}
    \label{fig:evap_scan}
\end{figure}

\begin{figure}[t]
    \centering
    \includegraphics[width=\textwidth]{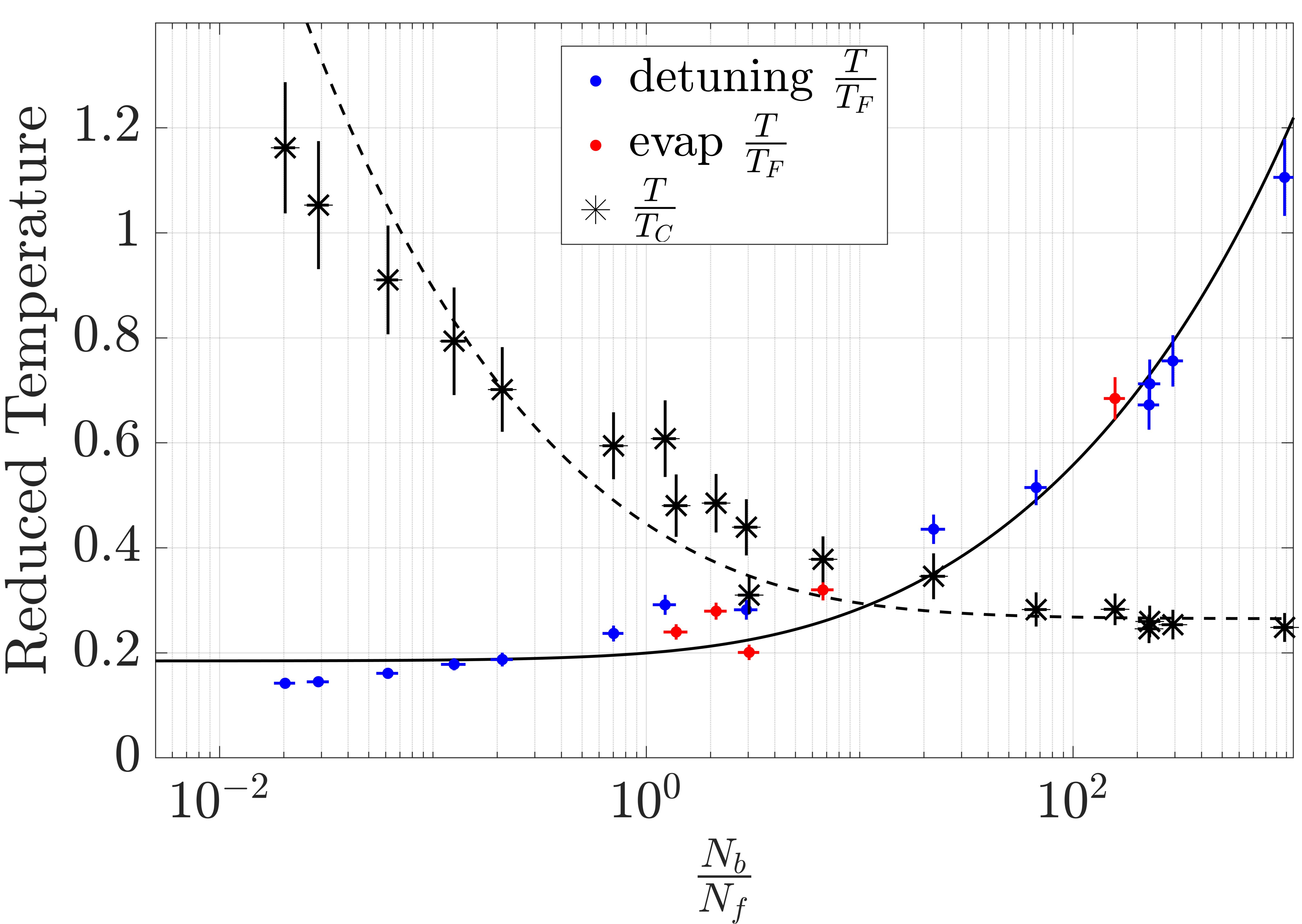}
    \caption{The reduced temperatures \(T/T_F\) and \(T/T_C\) plotted against the ratio of the number of bosons to fermions after the evaporation cycle. We see that \(T/T_F\) and \(N_b/N_f\) are equivalently related (solid line) for small changes to either the initial number of fermions (blue points) or evaporator height (red points) \cite{evap_note}. Using the relation \(T/T_C = 2.23 \left(\frac{N_b}{N_f}\right)^{-1/3} T/T_F\) we plot the \(T/T_C\) dependence on \(N_b/N_f\) (dashed line). The minimum of the average reduced temperature of both species is achieved for \(T/T_C=0.31(4)\) and \(T/T_F=0.20(3)\) with \(N_b/N_f=3.0(4)\).}
    \label{fig:T_ratio_vs_N_ratio}
\end{figure}

As the same relative detuning along the shared optical path is employed for both \HeThreeStar{} and \HeFourStar{} for all cooling we are effectively only able to change the initial amount of \HeThreeStar{} in our final magnetic trap via three methods: the ratio of the amount of \HeThree{} and \HeFour{} gas in the metastable source; the ratio of power between \HeThree{} and \HeFour{} light (controlled by changing the ratio of power in the seed light); and the detuning of the \HeThree{} seed light from its cooling transition, which can be controlled independently of the detuning of the \HeFour{} seed light. For this work we only focus on the behavior of \HeThree{} detuning as it is the only parameter of the three that changes the initial amount of \HeThreeStar{} atoms while leaving the initial amount and temperature of \HeFourStar{} atoms completely unchanged. 
All relative detuning and alignment of optomechanical components are optimized for \HeFourStar{} flux. As will be shown this is reasonable, as even with suboptimal conditions for \HeThreeStar{} we can still load more \HeThreeStar{} into the magnetic trap than we can effectively cool \cite{PhysRevA.69.043611}. In Fig.~\ref{fig:detuning_scan} we show how the atom numbers, temperature, and reduced temperatures vary with \HeThree{} detuning ($f_{ ^3He}-f_{ ^4He}$) for a fixed evaporator height of \(15\)~kHz. There is a region around zero detuning ($\Delta_{C3}=0$) where all \HeFourStar{} atoms are evaporated. We do not present a temperature measurement in this region due to the unreliability of the temperature measurement from the fermionic cloud. In this data set we have a minimum boson reduced temperature of \(T/T_C=0.25(1)\). Note that our system is capable of achieving significantly lower reduced temperatures of the boson cloud, however here we deliberately keep a significant portion of the \HeFourStar{} cloud thermal in order to more easily measure the temperature.

In Fig.~\ref{fig:evap_scan} we investigate the effect of evaporator height on the mixture's reduced temperatures and atom number with a fixed \HeThreeStar{} detuning of \(\Delta_{C3}=0.75\)~MHz. As expected, the mixtures, temperature and \HeFourStar{} atom number decrease with evaporator height, while the \HeThreeStar{} atom number remains roughly constant until almost all \HeFourStar{} atoms have been evaporated. The fermi reduced temperature hence decreases with evaporator height, while the bose reduced temperature increases when there is a non-zero number of \HeThreeStar{} atoms.

As discussed in Sec.~\ref{sec:evap}, the behavior of the reduced temperatures seen in Figs.~\ref{fig:detuning_scan} and \ref{fig:evap_scan} can be understood purely by how each variable (detuning of \HeThreeStar{} laser and final evaporator height) affect the ratio \(N_b/N_f\). In Fig.~\ref{fig:T_ratio_vs_N_ratio} we show how the Bose and Fermi reduced temperatures \(T/T_C\) and \(T/T_F\) vary with the atom number ratio \(N_b/N_f\). We find that the Fermi reduced temperature follows the expected relation derived from comparing heat capacities of the coolant (\HeFourStar{}) to the atom being sympathetically cooled (\HeThreeStar{}), and modeling the forced evaporation as a temperature dependent energy cutoff \cite{PhysRevA.69.043611,evap_note}. For our system the Bose and Fermi critical temperatures are related purely via the atom number ratio (Eqn.~\ref{eqn:TC_TF}). Thus, we can convert the relation between the Fermi reduced temperature and the atom number ratio to a relation for the bose reduced temperature.

Note that this relation holds regardless of the method via which these parameters are changed. This implies an optimal fermi reduced temperature for our current experimental conditions of \(T/T_F=0.18(3)\), in the limit of complete evaporation of \HeFourStar{}. This compares well to the minimum experimentally observed reduced temperature of \(T/T_F=0.14(1)\), achieved for an evaporator height of \(15\)~kHz and \HeThreeStar{} laser detuning of \(\Delta_{C3}=-1.6\)~MHz. We note that this value is well into the quantum degenerate regime.

\section{Conclusion}

In this work, we have successfully produced a degenerate Fermi gas of \HeThreeStar{}, with a minimum reduced temperature of \(T/T_F=0.14(1)\), via sympathetic cooling with \HeFourStar{} that is undergoing forced evaporation. The limiting reduced temperature could be improved further by using dynamic control of the relative intensity of the two frequencies of light, such that it is optimized for each stage of the experiment. This is especially relevant for the 1D Doppler cooling employed to initially cool both species, as this stage has a very strong effect on the initial phase space density of the mixture. The minimum reduced temperature can also be improved simply by increasing the heat capacity of the initial \HeFourStar{} sample.

The technique presented creates an excellent platform for a range of possible experiments in the areas of many-body physics, quantum atom optics, and Bose-Fermi mixtures. The capability of single particle detection for \HeThreeStar{} provides the opportunity to measure many-body correlations that display higher order antibunching, which has so far been unobserved. This behaviour is the equivalent Hanbury Brown-Twiss effect for fermions, which has been used to experimentally observe higher order bunching in bosonic \HeFourStar{} atoms \cite{Manning_2013,Dall2013}, and second order (two-body) antibunching in fermionic \HeThreeStar{} \cite{Jeltes2007}. It also allows access to interesting many body phenomena, such as creating an \(s\)-wave scattering halo \cite{PhysRevLett.118.240402} between a highly degenerate BEC and DFG, which will have a non-trivial range of Bose bunching and Fermi antibunching across the different correlation functions.

Furthermore, this \(s\)-wave scattering can be used to obtain a momentum entangled state of \HeThreeStar{} and \HeFourStar{}.  
Such a state could be used to test the validity of the weak equivalence principle for a nonclassical state \cite{PhysRevLett.120.043602}. While the weak equivalence principle has been extensively tested in the classical regime, its validity for quantum entities is still an open question.

Finally, this system could also be used to continue our spectroscopic investigation of the atomic structure of helium, for example the measurement of the \HeThreeStar{}-\HeFourStar{} \(s\)-wave scattering length \(a_{34}\) experimentally for the first time, as well as allowing us to observe thus far unseen quantum electrodynamic effects \cite{vanRooij196,Rengelink2018,doi:10.1126/science.abk2502}.

\bibliography{bibliography.bib}

\begin{thebibliography}{64}%
\makeatletter
\providecommand \@ifxundefined [1]{%
 \@ifx{#1\undefined}
}%
\providecommand \@ifnum [1]{%
 \ifnum #1\expandafter \@firstoftwo
 \else \expandafter \@secondoftwo
 \fi
}%
\providecommand \@ifx [1]{%
 \ifx #1\expandafter \@firstoftwo
 \else \expandafter \@secondoftwo
 \fi
}%
\providecommand \natexlab [1]{#1}%
\providecommand \enquote  [1]{``#1''}%
\providecommand \bibnamefont  [1]{#1}%
\providecommand \bibfnamefont [1]{#1}%
\providecommand \citenamefont [1]{#1}%
\providecommand \href@noop [0]{\@secondoftwo}%
\providecommand \href [0]{\begingroup \@sanitize@url \@href}%
\providecommand \@href[1]{\@@startlink{#1}\@@href}%
\providecommand \@@href[1]{\endgroup#1\@@endlink}%
\providecommand \@sanitize@url [0]{\catcode `\\12\catcode `\$12\catcode
  `\&12\catcode `\#12\catcode `\^12\catcode `\_12\catcode `\%12\relax}%
\providecommand \@@startlink[1]{}%
\providecommand \@@endlink[0]{}%
\providecommand \url  [0]{\begingroup\@sanitize@url \@url }%
\providecommand \@url [1]{\endgroup\@href {#1}{\urlprefix }}%
\providecommand \urlprefix  [0]{URL }%
\providecommand \Eprint [0]{\href }%
\providecommand \doibase [0]{https://doi.org/}%
\providecommand \selectlanguage [0]{\@gobble}%
\providecommand \bibinfo  [0]{\@secondoftwo}%
\providecommand \bibfield  [0]{\@secondoftwo}%
\providecommand \translation [1]{[#1]}%
\providecommand \BibitemOpen [0]{}%
\providecommand \bibitemStop [0]{}%
\providecommand \bibitemNoStop [0]{.\EOS\space}%
\providecommand \EOS [0]{\spacefactor3000\relax}%
\providecommand \BibitemShut  [1]{\csname bibitem#1\endcsname}%
\let\auto@bib@innerbib\@empty
\bibitem [{\citenamefont {Anderson}\ \emph {et~al.}(1995)\citenamefont
  {Anderson}, \citenamefont {Ensher}, \citenamefont {Matthews}, \citenamefont
  {Wieman},\ and\ \citenamefont {Cornell}}]{anderson1995observation}%
  \BibitemOpen
  \bibfield  {author} {\bibinfo {author} {\bibfnamefont {M.~H.}\ \bibnamefont
  {Anderson}}, \bibinfo {author} {\bibfnamefont {J.~R.}\ \bibnamefont
  {Ensher}}, \bibinfo {author} {\bibfnamefont {M.~R.}\ \bibnamefont
  {Matthews}}, \bibinfo {author} {\bibfnamefont {C.~E.}\ \bibnamefont
  {Wieman}},\ and\ \bibinfo {author} {\bibfnamefont {E.~A.}\ \bibnamefont
  {Cornell}},\ }\bibfield  {title} {\bibinfo {title} {Observation of
  {B}ose-{E}instein condensation in a dilute atomic vapor},\ }\href@noop {}
  {\bibfield  {journal} {\bibinfo  {journal} {science}\ }\textbf {\bibinfo
  {volume} {269}},\ \bibinfo {pages} {198} (\bibinfo {year}
  {1995})}\BibitemShut {NoStop}%
\bibitem [{\citenamefont {Jeltes}\ \emph {et~al.}(2007)\citenamefont {Jeltes},
  \citenamefont {McNamara}, \citenamefont {Hogervorst}, \citenamefont {Vassen},
  \citenamefont {Krachmalnicoff}, \citenamefont {Schellekens}, \citenamefont
  {Perrin}, \citenamefont {Chang}, \citenamefont {Boiron}, \citenamefont
  {Aspect},\ and\ \citenamefont {Westbrook}}]{Jeltes2007}%
  \BibitemOpen
  \bibfield  {author} {\bibinfo {author} {\bibfnamefont {T.}~\bibnamefont
  {Jeltes}}, \bibinfo {author} {\bibfnamefont {J.~M.}\ \bibnamefont
  {McNamara}}, \bibinfo {author} {\bibfnamefont {W.}~\bibnamefont
  {Hogervorst}}, \bibinfo {author} {\bibfnamefont {W.}~\bibnamefont {Vassen}},
  \bibinfo {author} {\bibfnamefont {V.}~\bibnamefont {Krachmalnicoff}},
  \bibinfo {author} {\bibfnamefont {M.}~\bibnamefont {Schellekens}}, \bibinfo
  {author} {\bibfnamefont {A.}~\bibnamefont {Perrin}}, \bibinfo {author}
  {\bibfnamefont {H.}~\bibnamefont {Chang}}, \bibinfo {author} {\bibfnamefont
  {D.}~\bibnamefont {Boiron}}, \bibinfo {author} {\bibfnamefont
  {A.}~\bibnamefont {Aspect}},\ and\ \bibinfo {author} {\bibfnamefont {C.~I.}\
  \bibnamefont {Westbrook}},\ }\bibfield  {title} {\bibinfo {title} {Comparison
  of the {H}anbury {B}rown-{T}wiss effect for bosons and fermions},\ }\href
  {https://doi.org/10.1038/nature05513} {\bibfield  {journal} {\bibinfo
  {journal} {Nature}\ }\textbf {\bibinfo {volume} {445}},\ \bibinfo {pages}
  {402} (\bibinfo {year} {2007})}\BibitemShut {NoStop}%
\bibitem [{\citenamefont {Henson}\ \emph
  {et~al.}(2022{\natexlab{a}})\citenamefont {Henson}, \citenamefont {Ross},
  \citenamefont {Thomas}, \citenamefont {Kuhn}, \citenamefont {Shin},
  \citenamefont {Hodgman}, \citenamefont {Zhang}, \citenamefont {Tang},
  \citenamefont {Drake}, \citenamefont {Bondy}, \citenamefont {Truscott},\ and\
  \citenamefont {Baldwin}}]{doi:10.1126/science.abk2502}%
  \BibitemOpen
  \bibfield  {author} {\bibinfo {author} {\bibfnamefont {B.~M.}\ \bibnamefont
  {Henson}}, \bibinfo {author} {\bibfnamefont {J.~A.}\ \bibnamefont {Ross}},
  \bibinfo {author} {\bibfnamefont {K.~F.}\ \bibnamefont {Thomas}}, \bibinfo
  {author} {\bibfnamefont {C.~N.}\ \bibnamefont {Kuhn}}, \bibinfo {author}
  {\bibfnamefont {D.~K.}\ \bibnamefont {Shin}}, \bibinfo {author}
  {\bibfnamefont {S.~S.}\ \bibnamefont {Hodgman}}, \bibinfo {author}
  {\bibfnamefont {Y.-H.}\ \bibnamefont {Zhang}}, \bibinfo {author}
  {\bibfnamefont {L.-Y.}\ \bibnamefont {Tang}}, \bibinfo {author}
  {\bibfnamefont {G.~W.~F.}\ \bibnamefont {Drake}}, \bibinfo {author}
  {\bibfnamefont {A.~T.}\ \bibnamefont {Bondy}}, \bibinfo {author}
  {\bibfnamefont {A.~G.}\ \bibnamefont {Truscott}},\ and\ \bibinfo {author}
  {\bibfnamefont {K.~G.~H.}\ \bibnamefont {Baldwin}},\ }\bibfield  {title}
  {\bibinfo {title} {Measurement of a helium tune-out frequency: an independent
  test of quantum electrodynamics},\ }\href
  {https://doi.org/10.1126/science.abk2502} {\bibfield  {journal} {\bibinfo
  {journal} {Science}\ }\textbf {\bibinfo {volume} {376}},\ \bibinfo {pages}
  {199} (\bibinfo {year} {2022}{\natexlab{a}})}\BibitemShut {NoStop}%
\bibitem [{\citenamefont {Byrnes}\ \emph {et~al.}(2012)\citenamefont {Byrnes},
  \citenamefont {Wen},\ and\ \citenamefont {Yamamoto}}]{PhysRevA.85.040306}%
  \BibitemOpen
  \bibfield  {author} {\bibinfo {author} {\bibfnamefont {T.}~\bibnamefont
  {Byrnes}}, \bibinfo {author} {\bibfnamefont {K.}~\bibnamefont {Wen}},\ and\
  \bibinfo {author} {\bibfnamefont {Y.}~\bibnamefont {Yamamoto}},\ }\bibfield
  {title} {\bibinfo {title} {Macroscopic quantum computation using
  {B}ose-{E}instein condensates},\ }\href
  {https://doi.org/10.1103/PhysRevA.85.040306} {\bibfield  {journal} {\bibinfo
  {journal} {Phys. Rev. A}\ }\textbf {\bibinfo {volume} {85}},\ \bibinfo
  {pages} {040306} (\bibinfo {year} {2012})}\BibitemShut {NoStop}%
\bibitem [{\citenamefont {Byrnes}\ \emph {et~al.}(2015)\citenamefont {Byrnes},
  \citenamefont {Rosseau}, \citenamefont {Khosla}, \citenamefont {Pyrkov},
  \citenamefont {Thomasen}, \citenamefont {Mukai}, \citenamefont {Koyama},
  \citenamefont {Abdelrahman},\ and\ \citenamefont
  {Ilo-Okeke}}]{BYRNES2015102}%
  \BibitemOpen
  \bibfield  {author} {\bibinfo {author} {\bibfnamefont {T.}~\bibnamefont
  {Byrnes}}, \bibinfo {author} {\bibfnamefont {D.}~\bibnamefont {Rosseau}},
  \bibinfo {author} {\bibfnamefont {M.}~\bibnamefont {Khosla}}, \bibinfo
  {author} {\bibfnamefont {A.}~\bibnamefont {Pyrkov}}, \bibinfo {author}
  {\bibfnamefont {A.}~\bibnamefont {Thomasen}}, \bibinfo {author}
  {\bibfnamefont {T.}~\bibnamefont {Mukai}}, \bibinfo {author} {\bibfnamefont
  {S.}~\bibnamefont {Koyama}}, \bibinfo {author} {\bibfnamefont
  {A.}~\bibnamefont {Abdelrahman}},\ and\ \bibinfo {author} {\bibfnamefont
  {E.}~\bibnamefont {Ilo-Okeke}},\ }\bibfield  {title} {\bibinfo {title}
  {Macroscopic quantum information processing using spin coherent states},\
  }\href {https://doi.org/https://doi.org/10.1016/j.optcom.2014.08.017}
  {\bibfield  {journal} {\bibinfo  {journal} {Optics Communications}\ }\textbf
  {\bibinfo {volume} {337}},\ \bibinfo {pages} {102} (\bibinfo {year}
  {2015})},\ \bibinfo {note} {macroscopic quantumness: theory and applications
  in optical sciences}\BibitemShut {NoStop}%
\bibitem [{\citenamefont {Murthy}\ \emph {et~al.}(2018)\citenamefont {Murthy},
  \citenamefont {Neidig}, \citenamefont {Klemt}, \citenamefont {Bayha},
  \citenamefont {Boettcher}, \citenamefont {Enss}, \citenamefont {Holten},
  \citenamefont {Z{\"u}rn}, \citenamefont {Preiss},\ and\ \citenamefont
  {Jochim}}]{murthy2018high}%
  \BibitemOpen
  \bibfield  {author} {\bibinfo {author} {\bibfnamefont {P.~A.}\ \bibnamefont
  {Murthy}}, \bibinfo {author} {\bibfnamefont {M.}~\bibnamefont {Neidig}},
  \bibinfo {author} {\bibfnamefont {R.}~\bibnamefont {Klemt}}, \bibinfo
  {author} {\bibfnamefont {L.}~\bibnamefont {Bayha}}, \bibinfo {author}
  {\bibfnamefont {I.}~\bibnamefont {Boettcher}}, \bibinfo {author}
  {\bibfnamefont {T.}~\bibnamefont {Enss}}, \bibinfo {author} {\bibfnamefont
  {M.}~\bibnamefont {Holten}}, \bibinfo {author} {\bibfnamefont
  {G.}~\bibnamefont {Z{\"u}rn}}, \bibinfo {author} {\bibfnamefont {P.~M.}\
  \bibnamefont {Preiss}},\ and\ \bibinfo {author} {\bibfnamefont
  {S.}~\bibnamefont {Jochim}},\ }\bibfield  {title} {\bibinfo {title}
  {High-temperature pairing in a strongly interacting two-dimensional {F}ermi
  gas},\ }\href@noop {} {\bibfield  {journal} {\bibinfo  {journal} {Science}\
  }\textbf {\bibinfo {volume} {359}},\ \bibinfo {pages} {452} (\bibinfo {year}
  {2018})}\BibitemShut {NoStop}%
\bibitem [{\citenamefont {Zwierlein}\ \emph {et~al.}(2005)\citenamefont
  {Zwierlein}, \citenamefont {Abo-Shaeer}, \citenamefont {Schirotzek},
  \citenamefont {Schunck},\ and\ \citenamefont
  {Ketterle}}]{zwierlein2005vortices}%
  \BibitemOpen
  \bibfield  {author} {\bibinfo {author} {\bibfnamefont {M.~W.}\ \bibnamefont
  {Zwierlein}}, \bibinfo {author} {\bibfnamefont {J.~R.}\ \bibnamefont
  {Abo-Shaeer}}, \bibinfo {author} {\bibfnamefont {A.}~\bibnamefont
  {Schirotzek}}, \bibinfo {author} {\bibfnamefont {C.~H.}\ \bibnamefont
  {Schunck}},\ and\ \bibinfo {author} {\bibfnamefont {W.}~\bibnamefont
  {Ketterle}},\ }\bibfield  {title} {\bibinfo {title} {Vortices and
  superfluidity in a strongly interacting {F}ermi gas},\ }\href@noop {}
  {\bibfield  {journal} {\bibinfo  {journal} {Nature}\ }\textbf {\bibinfo
  {volume} {435}},\ \bibinfo {pages} {1047} (\bibinfo {year}
  {2005})}\BibitemShut {NoStop}%
\bibitem [{\citenamefont {Chin}\ \emph {et~al.}(2006)\citenamefont {Chin},
  \citenamefont {Miller}, \citenamefont {Liu}, \citenamefont {Stan},
  \citenamefont {Setiawan}, \citenamefont {Sanner}, \citenamefont {Xu},\ and\
  \citenamefont {Ketterle}}]{chin2006evidence}%
  \BibitemOpen
  \bibfield  {author} {\bibinfo {author} {\bibfnamefont {J.~K.}\ \bibnamefont
  {Chin}}, \bibinfo {author} {\bibfnamefont {D.}~\bibnamefont {Miller}},
  \bibinfo {author} {\bibfnamefont {Y.}~\bibnamefont {Liu}}, \bibinfo {author}
  {\bibfnamefont {C.}~\bibnamefont {Stan}}, \bibinfo {author} {\bibfnamefont
  {W.}~\bibnamefont {Setiawan}}, \bibinfo {author} {\bibfnamefont
  {C.}~\bibnamefont {Sanner}}, \bibinfo {author} {\bibfnamefont
  {K.}~\bibnamefont {Xu}},\ and\ \bibinfo {author} {\bibfnamefont
  {W.}~\bibnamefont {Ketterle}},\ }\bibfield  {title} {\bibinfo {title}
  {Evidence for superfluidity of ultracold fermions in an optical lattice},\
  }\href@noop {} {\bibfield  {journal} {\bibinfo  {journal} {Nature}\ }\textbf
  {\bibinfo {volume} {443}},\ \bibinfo {pages} {961} (\bibinfo {year}
  {2006})}\BibitemShut {NoStop}%
\bibitem [{\citenamefont {Bloch}\ \emph {et~al.}(2008)\citenamefont {Bloch},
  \citenamefont {Dalibard},\ and\ \citenamefont {Zwerger}}]{RevModPhys.80.885}%
  \BibitemOpen
  \bibfield  {author} {\bibinfo {author} {\bibfnamefont {I.}~\bibnamefont
  {Bloch}}, \bibinfo {author} {\bibfnamefont {J.}~\bibnamefont {Dalibard}},\
  and\ \bibinfo {author} {\bibfnamefont {W.}~\bibnamefont {Zwerger}},\
  }\bibfield  {title} {\bibinfo {title} {Many-body physics with ultracold
  gases},\ }\href {https://doi.org/10.1103/RevModPhys.80.885} {\bibfield
  {journal} {\bibinfo  {journal} {Rev. Mod. Phys.}\ }\textbf {\bibinfo {volume}
  {80}},\ \bibinfo {pages} {885} (\bibinfo {year} {2008})}\BibitemShut
  {NoStop}%
\bibitem [{\citenamefont {Giorgini}\ \emph {et~al.}(2008)\citenamefont
  {Giorgini}, \citenamefont {Pitaevskii},\ and\ \citenamefont
  {Stringari}}]{RevModPhys.80.1215}%
  \BibitemOpen
  \bibfield  {author} {\bibinfo {author} {\bibfnamefont {S.}~\bibnamefont
  {Giorgini}}, \bibinfo {author} {\bibfnamefont {L.~P.}\ \bibnamefont
  {Pitaevskii}},\ and\ \bibinfo {author} {\bibfnamefont {S.}~\bibnamefont
  {Stringari}},\ }\bibfield  {title} {\bibinfo {title} {Theory of ultracold
  atomic {F}ermi gases},\ }\href {https://doi.org/10.1103/RevModPhys.80.1215}
  {\bibfield  {journal} {\bibinfo  {journal} {Rev. Mod. Phys.}\ }\textbf
  {\bibinfo {volume} {80}},\ \bibinfo {pages} {1215} (\bibinfo {year}
  {2008})}\BibitemShut {NoStop}%
\bibitem [{\citenamefont {Marchetti}\ \emph {et~al.}(2008)\citenamefont
  {Marchetti}, \citenamefont {Mathy}, \citenamefont {Huse},\ and\ \citenamefont
  {Parish}}]{PhysRevB.78.134517}%
  \BibitemOpen
  \bibfield  {author} {\bibinfo {author} {\bibfnamefont {F.~M.}\ \bibnamefont
  {Marchetti}}, \bibinfo {author} {\bibfnamefont {C.~J.~M.}\ \bibnamefont
  {Mathy}}, \bibinfo {author} {\bibfnamefont {D.~A.}\ \bibnamefont {Huse}},\
  and\ \bibinfo {author} {\bibfnamefont {M.~M.}\ \bibnamefont {Parish}},\
  }\bibfield  {title} {\bibinfo {title} {Phase separation and collapse in
  {B}ose-{F}ermi mixtures with a {F}eshbach resonance},\ }\href
  {https://doi.org/10.1103/PhysRevB.78.134517} {\bibfield  {journal} {\bibinfo
  {journal} {Phys. Rev. B}\ }\textbf {\bibinfo {volume} {78}},\ \bibinfo
  {pages} {134517} (\bibinfo {year} {2008})}\BibitemShut {NoStop}%
\bibitem [{\citenamefont {Bhongale}\ and\ \citenamefont
  {Pu}(2008)}]{PhysRevA.78.061606}%
  \BibitemOpen
  \bibfield  {author} {\bibinfo {author} {\bibfnamefont {S.~G.}\ \bibnamefont
  {Bhongale}}\ and\ \bibinfo {author} {\bibfnamefont {H.}~\bibnamefont {Pu}},\
  }\bibfield  {title} {\bibinfo {title} {{Phase separation in a mixture of a
  Bose-Einstein condensate and a two-component Fermi gas as a probe of Fermi
  superfluidity}},\ }\href {https://doi.org/10.1103/PhysRevA.78.061606}
  {\bibfield  {journal} {\bibinfo  {journal} {Phys. Rev. A}\ }\textbf {\bibinfo
  {volume} {78}},\ \bibinfo {pages} {061606} (\bibinfo {year}
  {2008})}\BibitemShut {NoStop}%
\bibitem [{\citenamefont {Linder}\ and\ \citenamefont
  {Sudb\o{}}(2010)}]{PhysRevA.81.013622}%
  \BibitemOpen
  \bibfield  {author} {\bibinfo {author} {\bibfnamefont {J.}~\bibnamefont
  {Linder}}\ and\ \bibinfo {author} {\bibfnamefont {A.}~\bibnamefont
  {Sudb\o{}}},\ }\bibfield  {title} {\bibinfo {title} {Probing phase separation
  in {B}ose-{F}ermi mixtures by the critical superfluid velocity},\ }\href
  {https://doi.org/10.1103/PhysRevA.81.013622} {\bibfield  {journal} {\bibinfo
  {journal} {Phys. Rev. A}\ }\textbf {\bibinfo {volume} {81}},\ \bibinfo
  {pages} {013622} (\bibinfo {year} {2010})}\BibitemShut {NoStop}%
\bibitem [{\citenamefont {Wu}\ \emph {et~al.}(2011)\citenamefont {Wu},
  \citenamefont {Santiago}, \citenamefont {Park}, \citenamefont {Ahmadi},\ and\
  \citenamefont {Zwierlein}}]{PhysRevA.84.011601}%
  \BibitemOpen
  \bibfield  {author} {\bibinfo {author} {\bibfnamefont {C.-H.}\ \bibnamefont
  {Wu}}, \bibinfo {author} {\bibfnamefont {I.}~\bibnamefont {Santiago}},
  \bibinfo {author} {\bibfnamefont {J.~W.}\ \bibnamefont {Park}}, \bibinfo
  {author} {\bibfnamefont {P.}~\bibnamefont {Ahmadi}},\ and\ \bibinfo {author}
  {\bibfnamefont {M.~W.}\ \bibnamefont {Zwierlein}},\ }\bibfield  {title}
  {\bibinfo {title} {Strongly interacting isotopic {B}ose-{F}ermi mixture
  immersed in a {F}ermi sea},\ }\href
  {https://doi.org/10.1103/PhysRevA.84.011601} {\bibfield  {journal} {\bibinfo
  {journal} {Phys. Rev. A}\ }\textbf {\bibinfo {volume} {84}},\ \bibinfo
  {pages} {011601} (\bibinfo {year} {2011})}\BibitemShut {NoStop}%
\bibitem [{\citenamefont {Fritsche}\ \emph {et~al.}(2021)\citenamefont
  {Fritsche}, \citenamefont {Baroni}, \citenamefont {Dobler}, \citenamefont
  {Kirilov}, \citenamefont {Huang}, \citenamefont {Grimm}, \citenamefont
  {Bruun},\ and\ \citenamefont {Massignan}}]{PhysRevA.103.053314}%
  \BibitemOpen
  \bibfield  {author} {\bibinfo {author} {\bibfnamefont {I.}~\bibnamefont
  {Fritsche}}, \bibinfo {author} {\bibfnamefont {C.}~\bibnamefont {Baroni}},
  \bibinfo {author} {\bibfnamefont {E.}~\bibnamefont {Dobler}}, \bibinfo
  {author} {\bibfnamefont {E.}~\bibnamefont {Kirilov}}, \bibinfo {author}
  {\bibfnamefont {B.}~\bibnamefont {Huang}}, \bibinfo {author} {\bibfnamefont
  {R.}~\bibnamefont {Grimm}}, \bibinfo {author} {\bibfnamefont {G.~M.}\
  \bibnamefont {Bruun}},\ and\ \bibinfo {author} {\bibfnamefont
  {P.}~\bibnamefont {Massignan}},\ }\bibfield  {title} {\bibinfo {title}
  {{Stability and breakdown of Fermi polarons in a strongly interacting
  Fermi-Bose mixture}},\ }\href {https://doi.org/10.1103/PhysRevA.103.053314}
  {\bibfield  {journal} {\bibinfo  {journal} {Phys. Rev. A}\ }\textbf {\bibinfo
  {volume} {103}},\ \bibinfo {pages} {053314} (\bibinfo {year}
  {2021})}\BibitemShut {NoStop}%
\bibitem [{\citenamefont {DeMarco}\ and\ \citenamefont
  {Jin}(1999)}]{doi:10.1126/science.285.5434.1703}%
  \BibitemOpen
  \bibfield  {author} {\bibinfo {author} {\bibfnamefont {B.}~\bibnamefont
  {DeMarco}}\ and\ \bibinfo {author} {\bibfnamefont {D.~S.}\ \bibnamefont
  {Jin}},\ }\bibfield  {title} {\bibinfo {title} {Onset of {F}ermi degeneracy
  in a trapped atomic gas},\ }\href
  {https://doi.org/10.1126/science.285.5434.1703} {\bibfield  {journal}
  {\bibinfo  {journal} {Science}\ }\textbf {\bibinfo {volume} {285}},\ \bibinfo
  {pages} {1703} (\bibinfo {year} {1999})}\BibitemShut {NoStop}%
\bibitem [{\citenamefont {Truscott}\ \emph {et~al.}(2001)\citenamefont
  {Truscott}, \citenamefont {Strecker}, \citenamefont {McAlexander},
  \citenamefont {Partridge},\ and\ \citenamefont
  {Hulet}}]{doi:10.1126/science.1059318}%
  \BibitemOpen
  \bibfield  {author} {\bibinfo {author} {\bibfnamefont {A.~G.}\ \bibnamefont
  {Truscott}}, \bibinfo {author} {\bibfnamefont {K.~E.}\ \bibnamefont
  {Strecker}}, \bibinfo {author} {\bibfnamefont {W.~I.}\ \bibnamefont
  {McAlexander}}, \bibinfo {author} {\bibfnamefont {G.~B.}\ \bibnamefont
  {Partridge}},\ and\ \bibinfo {author} {\bibfnamefont {R.~G.}\ \bibnamefont
  {Hulet}},\ }\bibfield  {title} {\bibinfo {title} {Observation of {F}ermi
  pressure in a gas of trapped atoms},\ }\href
  {https://doi.org/10.1126/science.1059318} {\bibfield  {journal} {\bibinfo
  {journal} {Science}\ }\textbf {\bibinfo {volume} {291}},\ \bibinfo {pages}
  {2570} (\bibinfo {year} {2001})}\BibitemShut {NoStop}%
\bibitem [{\citenamefont {Onofrio}(2016)}]{Onofrio_2016}%
  \BibitemOpen
  \bibfield  {author} {\bibinfo {author} {\bibfnamefont {R.}~\bibnamefont
  {Onofrio}},\ }\bibfield  {title} {\bibinfo {title} {Cooling and thermometry
  of atomic {F}ermi gases},\ }\href
  {https://doi.org/10.3367/ufne.2016.07.037873} {\bibfield  {journal} {\bibinfo
   {journal} {Physics-Uspekhi}\ }\textbf {\bibinfo {volume} {59}},\ \bibinfo
  {pages} {1129} (\bibinfo {year} {2016})}\BibitemShut {NoStop}%
\bibitem [{\citenamefont {Roati}\ \emph {et~al.}(2002)\citenamefont {Roati},
  \citenamefont {Riboli}, \citenamefont {Modugno},\ and\ \citenamefont
  {Inguscio}}]{PhysRevLett.89.150403}%
  \BibitemOpen
  \bibfield  {author} {\bibinfo {author} {\bibfnamefont {G.}~\bibnamefont
  {Roati}}, \bibinfo {author} {\bibfnamefont {F.}~\bibnamefont {Riboli}},
  \bibinfo {author} {\bibfnamefont {G.}~\bibnamefont {Modugno}},\ and\ \bibinfo
  {author} {\bibfnamefont {M.}~\bibnamefont {Inguscio}},\ }\bibfield  {title}
  {\bibinfo {title} {Fermi-bose quantum degenerate
  $^{\mathrm{40}}\mathrm{K}\mathrm{\text{\ensuremath{-}}}^{\mathrm{87}}\mathrm{R}\mathrm{b}$
  mixture with attractive interaction},\ }\href
  {https://doi.org/10.1103/PhysRevLett.89.150403} {\bibfield  {journal}
  {\bibinfo  {journal} {Phys. Rev. Lett.}\ }\textbf {\bibinfo {volume} {89}},\
  \bibinfo {pages} {150403} (\bibinfo {year} {2002})}\BibitemShut {NoStop}%
\bibitem [{\citenamefont {G\"unter}\ \emph {et~al.}(2006)\citenamefont
  {G\"unter}, \citenamefont {St\"oferle}, \citenamefont {Moritz}, \citenamefont
  {K\"ohl},\ and\ \citenamefont {Esslinger}}]{PhysRevLett.96.180402}%
  \BibitemOpen
  \bibfield  {author} {\bibinfo {author} {\bibfnamefont {K.}~\bibnamefont
  {G\"unter}}, \bibinfo {author} {\bibfnamefont {T.}~\bibnamefont
  {St\"oferle}}, \bibinfo {author} {\bibfnamefont {H.}~\bibnamefont {Moritz}},
  \bibinfo {author} {\bibfnamefont {M.}~\bibnamefont {K\"ohl}},\ and\ \bibinfo
  {author} {\bibfnamefont {T.}~\bibnamefont {Esslinger}},\ }\bibfield  {title}
  {\bibinfo {title} {{B}ose-{F}ermi mixtures in a three-dimensional optical
  lattice},\ }\href {https://doi.org/10.1103/PhysRevLett.96.180402} {\bibfield
  {journal} {\bibinfo  {journal} {Phys. Rev. Lett.}\ }\textbf {\bibinfo
  {volume} {96}},\ \bibinfo {pages} {180402} (\bibinfo {year}
  {2006})}\BibitemShut {NoStop}%
\bibitem [{\citenamefont {McNamara}\ \emph {et~al.}(2006)\citenamefont
  {McNamara}, \citenamefont {Jeltes}, \citenamefont {Tychkov}, \citenamefont
  {Hogervorst},\ and\ \citenamefont {Vassen}}]{PhysRevLett.97.080404}%
  \BibitemOpen
  \bibfield  {author} {\bibinfo {author} {\bibfnamefont {J.~M.}\ \bibnamefont
  {McNamara}}, \bibinfo {author} {\bibfnamefont {T.}~\bibnamefont {Jeltes}},
  \bibinfo {author} {\bibfnamefont {A.~S.}\ \bibnamefont {Tychkov}}, \bibinfo
  {author} {\bibfnamefont {W.}~\bibnamefont {Hogervorst}},\ and\ \bibinfo
  {author} {\bibfnamefont {W.}~\bibnamefont {Vassen}},\ }\bibfield  {title}
  {\bibinfo {title} {Degenerate {B}ose-{F}ermi mixture of metastable atoms},\
  }\href {https://doi.org/10.1103/PhysRevLett.97.080404} {\bibfield  {journal}
  {\bibinfo  {journal} {Phys. Rev. Lett.}\ }\textbf {\bibinfo {volume} {97}},\
  \bibinfo {pages} {080404} (\bibinfo {year} {2006})}\BibitemShut {NoStop}%
\bibitem [{\citenamefont {Schreck}\ \emph {et~al.}(2001)\citenamefont
  {Schreck}, \citenamefont {Ferrari}, \citenamefont {Corwin}, \citenamefont
  {Cubizolles}, \citenamefont {Khaykovich}, \citenamefont {Mewes},\ and\
  \citenamefont {Salomon}}]{PhysRevA.64.011402}%
  \BibitemOpen
  \bibfield  {author} {\bibinfo {author} {\bibfnamefont {F.}~\bibnamefont
  {Schreck}}, \bibinfo {author} {\bibfnamefont {G.}~\bibnamefont {Ferrari}},
  \bibinfo {author} {\bibfnamefont {K.~L.}\ \bibnamefont {Corwin}}, \bibinfo
  {author} {\bibfnamefont {J.}~\bibnamefont {Cubizolles}}, \bibinfo {author}
  {\bibfnamefont {L.}~\bibnamefont {Khaykovich}}, \bibinfo {author}
  {\bibfnamefont {M.-O.}\ \bibnamefont {Mewes}},\ and\ \bibinfo {author}
  {\bibfnamefont {C.}~\bibnamefont {Salomon}},\ }\bibfield  {title} {\bibinfo
  {title} {Sympathetic cooling of bosonic and fermionic lithium gases towards
  quantum degeneracy},\ }\href {https://doi.org/10.1103/PhysRevA.64.011402}
  {\bibfield  {journal} {\bibinfo  {journal} {Phys. Rev. A}\ }\textbf {\bibinfo
  {volume} {64}},\ \bibinfo {pages} {011402} (\bibinfo {year}
  {2001})}\BibitemShut {NoStop}%
\bibitem [{\citenamefont {Manning}\ \emph {et~al.}(2010)\citenamefont
  {Manning}, \citenamefont {Hodgman}, \citenamefont {Dall}, \citenamefont
  {Johnsson},\ and\ \citenamefont {Truscott}}]{Manning:10}%
  \BibitemOpen
  \bibfield  {author} {\bibinfo {author} {\bibfnamefont {A.~G.}\ \bibnamefont
  {Manning}}, \bibinfo {author} {\bibfnamefont {S.~S.}\ \bibnamefont
  {Hodgman}}, \bibinfo {author} {\bibfnamefont {R.~G.}\ \bibnamefont {Dall}},
  \bibinfo {author} {\bibfnamefont {M.~T.}\ \bibnamefont {Johnsson}},\ and\
  \bibinfo {author} {\bibfnamefont {A.~G.}\ \bibnamefont {Truscott}},\
  }\bibfield  {title} {\bibinfo {title} {The {H}anbury {B}rown-{T}wiss effect
  in a pulsed atom laser},\ }\href {https://doi.org/10.1364/OE.18.018712}
  {\bibfield  {journal} {\bibinfo  {journal} {Opt. Express}\ }\textbf {\bibinfo
  {volume} {18}},\ \bibinfo {pages} {18712} (\bibinfo {year}
  {2010})}\BibitemShut {NoStop}%
\bibitem [{\citenamefont {Manning}\ \emph {et~al.}(2015)\citenamefont
  {Manning}, \citenamefont {Khakimov}, \citenamefont {Dall},\ and\
  \citenamefont {Truscott}}]{Manning2015}%
  \BibitemOpen
  \bibfield  {author} {\bibinfo {author} {\bibfnamefont {A.~G.}\ \bibnamefont
  {Manning}}, \bibinfo {author} {\bibfnamefont {R.~I.}\ \bibnamefont
  {Khakimov}}, \bibinfo {author} {\bibfnamefont {R.~G.}\ \bibnamefont {Dall}},\
  and\ \bibinfo {author} {\bibfnamefont {A.~G.}\ \bibnamefont {Truscott}},\
  }\bibfield  {title} {\bibinfo {title} {Wheeler's delayed-choice gedanken
  experiment with a single atom},\ }\href {https://doi.org/10.1038/nphys3343}
  {\bibfield  {journal} {\bibinfo  {journal} {Nature Physics}\ }\textbf
  {\bibinfo {volume} {11}},\ \bibinfo {pages} {539} (\bibinfo {year}
  {2015})}\BibitemShut {NoStop}%
\bibitem [{\citenamefont {Lopes}\ \emph {et~al.}(2015)\citenamefont {Lopes},
  \citenamefont {Imanaliev}, \citenamefont {Aspect}, \citenamefont {Cheneau},
  \citenamefont {Boiron},\ and\ \citenamefont {Westbrook}}]{Lopes2015}%
  \BibitemOpen
  \bibfield  {author} {\bibinfo {author} {\bibfnamefont {R.}~\bibnamefont
  {Lopes}}, \bibinfo {author} {\bibfnamefont {A.}~\bibnamefont {Imanaliev}},
  \bibinfo {author} {\bibfnamefont {A.}~\bibnamefont {Aspect}}, \bibinfo
  {author} {\bibfnamefont {M.}~\bibnamefont {Cheneau}}, \bibinfo {author}
  {\bibfnamefont {D.}~\bibnamefont {Boiron}},\ and\ \bibinfo {author}
  {\bibfnamefont {C.~I.}\ \bibnamefont {Westbrook}},\ }\bibfield  {title}
  {\bibinfo {title} {Atomic {H}ong--{O}u--{M}andel experiment},\ }\href
  {https://doi.org/10.1038/nature14331} {\bibfield  {journal} {\bibinfo
  {journal} {Nature}\ }\textbf {\bibinfo {volume} {520}},\ \bibinfo {pages}
  {66} (\bibinfo {year} {2015})}\BibitemShut {NoStop}%
\bibitem [{\citenamefont {Dussarrat}\ \emph {et~al.}(2017)\citenamefont
  {Dussarrat}, \citenamefont {Perrier}, \citenamefont {Imanaliev},
  \citenamefont {Lopes}, \citenamefont {Aspect}, \citenamefont {Cheneau},
  \citenamefont {Boiron},\ and\ \citenamefont
  {Westbrook}}]{PhysRevLett.119.173202}%
  \BibitemOpen
  \bibfield  {author} {\bibinfo {author} {\bibfnamefont {P.}~\bibnamefont
  {Dussarrat}}, \bibinfo {author} {\bibfnamefont {M.}~\bibnamefont {Perrier}},
  \bibinfo {author} {\bibfnamefont {A.}~\bibnamefont {Imanaliev}}, \bibinfo
  {author} {\bibfnamefont {R.}~\bibnamefont {Lopes}}, \bibinfo {author}
  {\bibfnamefont {A.}~\bibnamefont {Aspect}}, \bibinfo {author} {\bibfnamefont
  {M.}~\bibnamefont {Cheneau}}, \bibinfo {author} {\bibfnamefont
  {D.}~\bibnamefont {Boiron}},\ and\ \bibinfo {author} {\bibfnamefont {C.~I.}\
  \bibnamefont {Westbrook}},\ }\bibfield  {title} {\bibinfo {title}
  {{Two-Particle Four-Mode Interferometer for Atoms}},\ }\href
  {https://doi.org/10.1103/PhysRevLett.119.173202} {\bibfield  {journal}
  {\bibinfo  {journal} {Phys. Rev. Lett.}\ }\textbf {\bibinfo {volume} {119}},\
  \bibinfo {pages} {173202} (\bibinfo {year} {2017})}\BibitemShut {NoStop}%
\bibitem [{\citenamefont {van Rooij}\ \emph {et~al.}(2011)\citenamefont {van
  Rooij}, \citenamefont {Borbely}, \citenamefont {Simonet}, \citenamefont
  {Hoogerland}, \citenamefont {Eikema}, \citenamefont {Rozendaal},\ and\
  \citenamefont {Vassen}}]{vanRooij196}%
  \BibitemOpen
  \bibfield  {author} {\bibinfo {author} {\bibfnamefont {R.}~\bibnamefont {van
  Rooij}}, \bibinfo {author} {\bibfnamefont {J.~S.}\ \bibnamefont {Borbely}},
  \bibinfo {author} {\bibfnamefont {J.}~\bibnamefont {Simonet}}, \bibinfo
  {author} {\bibfnamefont {M.~D.}\ \bibnamefont {Hoogerland}}, \bibinfo
  {author} {\bibfnamefont {K.~S.~E.}\ \bibnamefont {Eikema}}, \bibinfo {author}
  {\bibfnamefont {R.~A.}\ \bibnamefont {Rozendaal}},\ and\ \bibinfo {author}
  {\bibfnamefont {W.}~\bibnamefont {Vassen}},\ }\bibfield  {title} {\bibinfo
  {title} {Frequency metrology in quantum degenerate helium: Direct measurement
  of the $2^{3}{S}_{1} \rightarrow 2^{1}{S}_{0}$ transition},\ }\href
  {https://doi.org/10.1126/science.1205163} {\bibfield  {journal} {\bibinfo
  {journal} {Science}\ }\textbf {\bibinfo {volume} {333}},\ \bibinfo {pages}
  {196} (\bibinfo {year} {2011})}\BibitemShut {NoStop}%
\bibitem [{\citenamefont {Rengelink}\ \emph {et~al.}(2018)\citenamefont
  {Rengelink}, \citenamefont {van~der Werf}, \citenamefont {Notermans},
  \citenamefont {Jannin}, \citenamefont {Eikema}, \citenamefont {Hoogerland},\
  and\ \citenamefont {Vassen}}]{Rengelink2018}%
  \BibitemOpen
  \bibfield  {author} {\bibinfo {author} {\bibfnamefont {R.~J.}\ \bibnamefont
  {Rengelink}}, \bibinfo {author} {\bibfnamefont {Y.}~\bibnamefont {van~der
  Werf}}, \bibinfo {author} {\bibfnamefont {R.~P. M. J.~W.}\ \bibnamefont
  {Notermans}}, \bibinfo {author} {\bibfnamefont {R.}~\bibnamefont {Jannin}},
  \bibinfo {author} {\bibfnamefont {K.~S.~E.}\ \bibnamefont {Eikema}}, \bibinfo
  {author} {\bibfnamefont {M.~D.}\ \bibnamefont {Hoogerland}},\ and\ \bibinfo
  {author} {\bibfnamefont {W.}~\bibnamefont {Vassen}},\ }\bibfield  {title}
  {\bibinfo {title} {Precision spectroscopy of helium in a magic wavelength
  optical dipole trap},\ }\href {https://doi.org/10.1038/s41567-018-0242-5}
  {\bibfield  {journal} {\bibinfo  {journal} {Nature Physics}\ }\textbf
  {\bibinfo {volume} {14}},\ \bibinfo {pages} {1132} (\bibinfo {year}
  {2018})}\BibitemShut {NoStop}%
\bibitem [{\citenamefont {Geiger}\ and\ \citenamefont
  {Trupke}(2018)}]{PhysRevLett.120.043602}%
  \BibitemOpen
  \bibfield  {author} {\bibinfo {author} {\bibfnamefont {R.}~\bibnamefont
  {Geiger}}\ and\ \bibinfo {author} {\bibfnamefont {M.}~\bibnamefont
  {Trupke}},\ }\bibfield  {title} {\bibinfo {title} {Proposal for a quantum
  test of the weak equivalence principle with entangled atomic species},\
  }\href {https://doi.org/10.1103/PhysRevLett.120.043602} {\bibfield  {journal}
  {\bibinfo  {journal} {Phys. Rev. Lett.}\ }\textbf {\bibinfo {volume} {120}},\
  \bibinfo {pages} {043602} (\bibinfo {year} {2018})}\BibitemShut {NoStop}%
\bibitem [{\citenamefont {Dall}\ and\ \citenamefont
  {Truscott}(2007)}]{Dall2007}%
  \BibitemOpen
  \bibfield  {author} {\bibinfo {author} {\bibfnamefont {R.}~\bibnamefont
  {Dall}}\ and\ \bibinfo {author} {\bibfnamefont {A.}~\bibnamefont
  {Truscott}},\ }\bibfield  {title} {\bibinfo {title}
  {Bose{\textendash}{E}instein condensation of metastable helium in a bi-planar
  quadrupole {I}offe configuration trap},\ }\href
  {https://doi.org/10.1016/j.optcom.2006.09.031} {\bibfield  {journal}
  {\bibinfo  {journal} {Opt. Commun.}\ }\textbf {\bibinfo {volume} {270}},\
  \bibinfo {pages} {255} (\bibinfo {year} {2007})}\BibitemShut {NoStop}%
\bibitem [{\citenamefont {Pethick}\ and\ \citenamefont
  {Smith}(2008)}]{pethick2008bose}%
  \BibitemOpen
  \bibfield  {author} {\bibinfo {author} {\bibfnamefont {C.~J.}\ \bibnamefont
  {Pethick}}\ and\ \bibinfo {author} {\bibfnamefont {H.}~\bibnamefont
  {Smith}},\ }\href@noop {} {\emph {\bibinfo {title} {Bose--Einstein
  condensation in dilute gases}}}\ (\bibinfo  {publisher} {Cambridge university
  press},\ \bibinfo {year} {2008})\BibitemShut {NoStop}%
\bibitem [{\citenamefont {Castin}\ and\ \citenamefont
  {Dum}(1996)}]{PhysRevLett.77.5315}%
  \BibitemOpen
  \bibfield  {author} {\bibinfo {author} {\bibfnamefont {Y.}~\bibnamefont
  {Castin}}\ and\ \bibinfo {author} {\bibfnamefont {R.}~\bibnamefont {Dum}},\
  }\bibfield  {title} {\bibinfo {title} {{Bose-Einstein Condensates in Time
  Dependent Traps}},\ }\href {https://doi.org/10.1103/PhysRevLett.77.5315}
  {\bibfield  {journal} {\bibinfo  {journal} {Phys. Rev. Lett.}\ }\textbf
  {\bibinfo {volume} {77}},\ \bibinfo {pages} {5315} (\bibinfo {year}
  {1996})}\BibitemShut {NoStop}%
\bibitem [{\citenamefont {Moal}\ \emph {et~al.}(2006)\citenamefont {Moal},
  \citenamefont {Portier}, \citenamefont {Kim}, \citenamefont {Dugu\'e},
  \citenamefont {Rapol}, \citenamefont {Leduc},\ and\ \citenamefont
  {Cohen-Tannoudji}}]{PhysRevLett.96.023203}%
  \BibitemOpen
  \bibfield  {author} {\bibinfo {author} {\bibfnamefont {S.}~\bibnamefont
  {Moal}}, \bibinfo {author} {\bibfnamefont {M.}~\bibnamefont {Portier}},
  \bibinfo {author} {\bibfnamefont {J.}~\bibnamefont {Kim}}, \bibinfo {author}
  {\bibfnamefont {J.}~\bibnamefont {Dugu\'e}}, \bibinfo {author} {\bibfnamefont
  {U.~D.}\ \bibnamefont {Rapol}}, \bibinfo {author} {\bibfnamefont
  {M.}~\bibnamefont {Leduc}},\ and\ \bibinfo {author} {\bibfnamefont
  {C.}~\bibnamefont {Cohen-Tannoudji}},\ }\bibfield  {title} {\bibinfo {title}
  {Accurate determination of the scattering length of metastable helium atoms
  using dark resonances between atoms and exotic molecules},\ }\href
  {https://doi.org/10.1103/PhysRevLett.96.023203} {\bibfield  {journal}
  {\bibinfo  {journal} {Phys. Rev. Lett.}\ }\textbf {\bibinfo {volume} {96}},\
  \bibinfo {pages} {023203} (\bibinfo {year} {2006})}\BibitemShut {NoStop}%
\bibitem [{\citenamefont {Baym}\ and\ \citenamefont
  {Pethick}(1996)}]{PhysRevLett.76.6}%
  \BibitemOpen
  \bibfield  {author} {\bibinfo {author} {\bibfnamefont {G.}~\bibnamefont
  {Baym}}\ and\ \bibinfo {author} {\bibfnamefont {C.~J.}\ \bibnamefont
  {Pethick}},\ }\bibfield  {title} {\bibinfo {title} {Ground-state properties
  of magnetically trapped {B}ose-condensed rubidium gas},\ }\href
  {https://doi.org/10.1103/PhysRevLett.76.6} {\bibfield  {journal} {\bibinfo
  {journal} {Phys. Rev. Lett.}\ }\textbf {\bibinfo {volume} {76}},\ \bibinfo
  {pages} {6} (\bibinfo {year} {1996})}\BibitemShut {NoStop}%
\bibitem [{\citenamefont {Butts}\ and\ \citenamefont
  {Rokhsar}(1997)}]{PhysRevA.55.4346}%
  \BibitemOpen
  \bibfield  {author} {\bibinfo {author} {\bibfnamefont {D.~A.}\ \bibnamefont
  {Butts}}\ and\ \bibinfo {author} {\bibfnamefont {D.~S.}\ \bibnamefont
  {Rokhsar}},\ }\bibfield  {title} {\bibinfo {title} {Trapped {F}ermi gases},\
  }\href {https://doi.org/10.1103/PhysRevA.55.4346} {\bibfield  {journal}
  {\bibinfo  {journal} {Phys. Rev. A}\ }\textbf {\bibinfo {volume} {55}},\
  \bibinfo {pages} {4346} (\bibinfo {year} {1997})}\BibitemShut {NoStop}%
\bibitem [{\citenamefont {DeMarco}(2001)}]{demarco2001quantum}%
  \BibitemOpen
  \bibfield  {author} {\bibinfo {author} {\bibfnamefont {B.~L.}\ \bibnamefont
  {DeMarco}},\ }\href@noop {} {\emph {\bibinfo {title} {Quantum behavior of an
  atomic Fermi gas}}}\ (\bibinfo  {publisher} {University of Colorado at
  Boulder},\ \bibinfo {year} {2001})\BibitemShut {NoStop}%
\bibitem [{\citenamefont {Yavin}\ \emph {et~al.}(2002)\citenamefont {Yavin},
  \citenamefont {Weel}, \citenamefont {Andreyuk},\ and\ \citenamefont
  {Kumarakrishnan}}]{Yavin2002}%
  \BibitemOpen
  \bibfield  {author} {\bibinfo {author} {\bibfnamefont {I.}~\bibnamefont
  {Yavin}}, \bibinfo {author} {\bibfnamefont {M.}~\bibnamefont {Weel}},
  \bibinfo {author} {\bibfnamefont {A.}~\bibnamefont {Andreyuk}},\ and\
  \bibinfo {author} {\bibfnamefont {A.}~\bibnamefont {Kumarakrishnan}},\
  }\bibfield  {title} {\bibinfo {title} {A calculation of the time-of-flight
  distribution of trapped atoms},\ }\href {https://doi.org/10.1119/1.1424266}
  {\bibfield  {journal} {\bibinfo  {journal} {American Journal of Physics}\
  }\textbf {\bibinfo {volume} {70}},\ \bibinfo {pages} {149} (\bibinfo {year}
  {2002})}\BibitemShut {NoStop}%
\bibitem [{dis()}]{dist_note}%
  \BibitemOpen
  \href@noop {} {}\bibinfo {note} {If we wish to fit a thermal distribution to
  the wings of our time-of-flight profiles it is relevant to know that for a
  given reduced temperature and momentum cuttoff (i.e. all data with momenta
  smaller than the cuttoff is ignored for purpose of the fitting) the fitted
  temperature is more strongly dependent on the fugacity for fermions, compared
  to bosons. There is a small factor which contributes to this specific to our
  experiment, namely \(v_b<v_f\) for a fixed temperature as \(m_b>m_f\), and
  thus the Bose distribution decays more rapidly. The primary factors are that
  the Fermi fugacity \(\xi_f\) grows significantly faster with respect to
  reduced temperature \(\tau\) compared to the Bose fugacity \(\xi_b\). This
  can be readily seen from the relations
  $\tau^{-3/2}=\Gamma(5/2)\text{Li}_{\frac{3}{2}} (\xi_b)$ and
  $\tau^{-3/2}=-\Gamma(5/2)\text{Li}_{\frac{3}{2}} (-\xi_f)$ which are valid
  for a harmonic trap \cite{cowan2019chemical}. Furthermore, Fermi fugacity has
  a stronger effect on the high momentum wings as it has a broadening effect on
  the distribution (compare to a purely thermal distribution). Conversely, the
  Bose distribution narrows with increasing fugacity and thus is less effected
  in the wings.}\BibitemShut {Stop}%
\bibitem [{\citenamefont {Hodgman}\ \emph
  {et~al.}(2009{\natexlab{a}})\citenamefont {Hodgman}, \citenamefont {Dall},
  \citenamefont {Byron}, \citenamefont {Baldwin}, \citenamefont {Buckman},\
  and\ \citenamefont {Truscott}}]{PhysRevLett.103.053002}%
  \BibitemOpen
  \bibfield  {author} {\bibinfo {author} {\bibfnamefont {S.~S.}\ \bibnamefont
  {Hodgman}}, \bibinfo {author} {\bibfnamefont {R.~G.}\ \bibnamefont {Dall}},
  \bibinfo {author} {\bibfnamefont {L.~J.}\ \bibnamefont {Byron}}, \bibinfo
  {author} {\bibfnamefont {K.~G.~H.}\ \bibnamefont {Baldwin}}, \bibinfo
  {author} {\bibfnamefont {S.~J.}\ \bibnamefont {Buckman}},\ and\ \bibinfo
  {author} {\bibfnamefont {A.~G.}\ \bibnamefont {Truscott}},\ }\bibfield
  {title} {\bibinfo {title} {Metastable helium: A new determination of the
  longest atomic excited-state lifetime},\ }\href
  {https://doi.org/10.1103/PhysRevLett.103.053002} {\bibfield  {journal}
  {\bibinfo  {journal} {Phys. Rev. Lett.}\ }\textbf {\bibinfo {volume} {103}},\
  \bibinfo {pages} {053002} (\bibinfo {year} {2009}{\natexlab{a}})}\BibitemShut
  {NoStop}%
\bibitem [{\citenamefont {Jeltes}(2008)}]{jeltesphdthesis}%
  \BibitemOpen
  \bibfield  {author} {\bibinfo {author} {\bibfnamefont {T.}~\bibnamefont
  {Jeltes}},\ }\emph {\bibinfo {title} {Quantum Statistical Effects in
  Ultracold Gases of Metastable Helium}},\ \href@noop {} {Ph.D. thesis},\
  \bibinfo  {school} {Vrije Universiteit Amsterdam} (\bibinfo {year}
  {2008})\BibitemShut {NoStop}%
\bibitem [{\citenamefont {Hodgman}\ \emph
  {et~al.}(2009{\natexlab{b}})\citenamefont {Hodgman}, \citenamefont {Dall},
  \citenamefont {Baldwin},\ and\ \citenamefont {Truscott}}]{Hodgman2009a}%
  \BibitemOpen
  \bibfield  {author} {\bibinfo {author} {\bibfnamefont {S.~S.}\ \bibnamefont
  {Hodgman}}, \bibinfo {author} {\bibfnamefont {R.~G.}\ \bibnamefont {Dall}},
  \bibinfo {author} {\bibfnamefont {K.~G.~H.}\ \bibnamefont {Baldwin}},\ and\
  \bibinfo {author} {\bibfnamefont {A.~G.}\ \bibnamefont {Truscott}},\
  }\bibfield  {title} {\bibinfo {title} {Complete ground-state transition rates
  for the helium $2\text{ }^{3}{P}$ manifold},\ }\href
  {https://doi.org/10.1103/PhysRevA.80.044501} {\bibfield  {journal} {\bibinfo
  {journal} {Phys. Rev. A}\ }\textbf {\bibinfo {volume} {80}},\ \bibinfo
  {pages} {044501} (\bibinfo {year} {2009}{\natexlab{b}})}\BibitemShut
  {NoStop}%
\bibitem [{\citenamefont {Prestage}\ \emph {et~al.}(1985)\citenamefont
  {Prestage}, \citenamefont {Johnson}, \citenamefont {Hinds},\ and\
  \citenamefont {Pichanick}}]{PhysRevA.32.2712}%
  \BibitemOpen
  \bibfield  {author} {\bibinfo {author} {\bibfnamefont {J.~D.}\ \bibnamefont
  {Prestage}}, \bibinfo {author} {\bibfnamefont {C.~E.}\ \bibnamefont
  {Johnson}}, \bibinfo {author} {\bibfnamefont {E.~A.}\ \bibnamefont {Hinds}},\
  and\ \bibinfo {author} {\bibfnamefont {F.~M.~J.}\ \bibnamefont {Pichanick}},\
  }\bibfield  {title} {\bibinfo {title} {Precise study of hyperfine structure
  in the ${2}^{3}$p state of $^{3}\mathrm{He}$},\ }\href
  {https://doi.org/10.1103/PhysRevA.32.2712} {\bibfield  {journal} {\bibinfo
  {journal} {Phys. Rev. A}\ }\textbf {\bibinfo {volume} {32}},\ \bibinfo
  {pages} {2712} (\bibinfo {year} {1985})}\BibitemShut {NoStop}%
\bibitem [{\citenamefont {Vassen}\ \emph {et~al.}(2012)\citenamefont {Vassen},
  \citenamefont {Cohen-Tannoudji}, \citenamefont {Leduc}, \citenamefont
  {Boiron}, \citenamefont {Westbrook}, \citenamefont {Truscott}, \citenamefont
  {Baldwin}, \citenamefont {Birkl}, \citenamefont {Cancio},\ and\ \citenamefont
  {Trippenbach}}]{RevModPhys.84.175}%
  \BibitemOpen
  \bibfield  {author} {\bibinfo {author} {\bibfnamefont {W.}~\bibnamefont
  {Vassen}}, \bibinfo {author} {\bibfnamefont {C.}~\bibnamefont
  {Cohen-Tannoudji}}, \bibinfo {author} {\bibfnamefont {M.}~\bibnamefont
  {Leduc}}, \bibinfo {author} {\bibfnamefont {D.}~\bibnamefont {Boiron}},
  \bibinfo {author} {\bibfnamefont {C.~I.}\ \bibnamefont {Westbrook}}, \bibinfo
  {author} {\bibfnamefont {A.}~\bibnamefont {Truscott}}, \bibinfo {author}
  {\bibfnamefont {K.}~\bibnamefont {Baldwin}}, \bibinfo {author} {\bibfnamefont
  {G.}~\bibnamefont {Birkl}}, \bibinfo {author} {\bibfnamefont
  {P.}~\bibnamefont {Cancio}},\ and\ \bibinfo {author} {\bibfnamefont
  {M.}~\bibnamefont {Trippenbach}},\ }\bibfield  {title} {\bibinfo {title}
  {Cold and trapped metastable noble gases},\ }\href
  {https://doi.org/10.1103/RevModPhys.84.175} {\bibfield  {journal} {\bibinfo
  {journal} {Rev. Mod. Phys.}\ }\textbf {\bibinfo {volume} {84}},\ \bibinfo
  {pages} {175} (\bibinfo {year} {2012})}\BibitemShut {NoStop}%
\bibitem [{\citenamefont {Stas}(2005)}]{Stas_phdthesis}%
  \BibitemOpen
  \bibfield  {author} {\bibinfo {author} {\bibfnamefont {R.~J.~W.}\
  \bibnamefont {Stas}},\ }\emph {\bibinfo {title} {Trapping fermionic and
  bosonic helium atoms}},\ \href@noop {} {Ph.D. thesis},\ \bibinfo  {school}
  {Vrije Universiteit Amsterdam} (\bibinfo {year} {2005})\BibitemShut {NoStop}%
\bibitem [{\citenamefont {Pastor}\ \emph {et~al.}(2004)\citenamefont {Pastor},
  \citenamefont {Giusfredi}, \citenamefont {Natale}, \citenamefont {Hagel},
  \citenamefont {de~Mauro},\ and\ \citenamefont
  {Inguscio}}]{PhysRevLett.92.023001}%
  \BibitemOpen
  \bibfield  {author} {\bibinfo {author} {\bibfnamefont {P.~C.}\ \bibnamefont
  {Pastor}}, \bibinfo {author} {\bibfnamefont {G.}~\bibnamefont {Giusfredi}},
  \bibinfo {author} {\bibfnamefont {P.~D.}\ \bibnamefont {Natale}}, \bibinfo
  {author} {\bibfnamefont {G.}~\bibnamefont {Hagel}}, \bibinfo {author}
  {\bibfnamefont {C.}~\bibnamefont {de~Mauro}},\ and\ \bibinfo {author}
  {\bibfnamefont {M.}~\bibnamefont {Inguscio}},\ }\bibfield  {title} {\bibinfo
  {title} {Absolute frequency measurements of the
  \ensuremath${2}^{3}{S}_{1}{\rightarrow}{2}^{3}{P}_{0,1,2}$ atomic helium
  transitions around 1083 nm},\ }\href
  {https://doi.org/10.1103/PhysRevLett.92.023001} {\bibfield  {journal}
  {\bibinfo  {journal} {Phys. Rev. Lett.}\ }\textbf {\bibinfo {volume} {92}},\
  \bibinfo {pages} {023001} (\bibinfo {year} {2004})}\BibitemShut {NoStop}%
\bibitem [{\citenamefont {Rosner}\ and\ \citenamefont
  {Pipkin}(1970)}]{PhysRevA.1.571}%
  \BibitemOpen
  \bibfield  {author} {\bibinfo {author} {\bibfnamefont {S.~D.}\ \bibnamefont
  {Rosner}}\ and\ \bibinfo {author} {\bibfnamefont {F.~M.}\ \bibnamefont
  {Pipkin}},\ }\bibfield  {title} {\bibinfo {title} {Hyperfine structure of the
  $2^{3}{S}_{1}$ state of ${\mathrm{{h}e}}^{3}$},\ }\href
  {https://doi.org/10.1103/PhysRevA.1.571} {\bibfield  {journal} {\bibinfo
  {journal} {Phys. Rev. A}\ }\textbf {\bibinfo {volume} {1}},\ \bibinfo {pages}
  {571} (\bibinfo {year} {1970})}\BibitemShut {NoStop}%
\bibitem [{\citenamefont {Geist}\ \emph {et~al.}(1999)\citenamefont {Geist},
  \citenamefont {Idrizbegovic}, \citenamefont {Marinescu}, \citenamefont
  {Kennedy},\ and\ \citenamefont {You}}]{PhysRevA.61.013406}%
  \BibitemOpen
  \bibfield  {author} {\bibinfo {author} {\bibfnamefont {W.}~\bibnamefont
  {Geist}}, \bibinfo {author} {\bibfnamefont {A.}~\bibnamefont {Idrizbegovic}},
  \bibinfo {author} {\bibfnamefont {M.}~\bibnamefont {Marinescu}}, \bibinfo
  {author} {\bibfnamefont {T.~A.~B.}\ \bibnamefont {Kennedy}},\ and\ \bibinfo
  {author} {\bibfnamefont {L.}~\bibnamefont {You}},\ }\bibfield  {title}
  {\bibinfo {title} {Evaporative cooling of trapped fermionic atoms},\ }\href
  {https://doi.org/10.1103/PhysRevA.61.013406} {\bibfield  {journal} {\bibinfo
  {journal} {Phys. Rev. A}\ }\textbf {\bibinfo {volume} {61}},\ \bibinfo
  {pages} {013406} (\bibinfo {year} {1999})}\BibitemShut {NoStop}%
\bibitem [{\citenamefont {Hirsch}\ \emph {et~al.}(2021)\citenamefont {Hirsch},
  \citenamefont {Cocks},\ and\ \citenamefont {Hodgman}}]{PhysRevA.104.033317}%
  \BibitemOpen
  \bibfield  {author} {\bibinfo {author} {\bibfnamefont {T.~M.~F.}\
  \bibnamefont {Hirsch}}, \bibinfo {author} {\bibfnamefont {D.~G.}\
  \bibnamefont {Cocks}},\ and\ \bibinfo {author} {\bibfnamefont {S.~S.}\
  \bibnamefont {Hodgman}},\ }\bibfield  {title} {\bibinfo {title}
  {Close-coupled model of {F}eshbach resonances in ultracold
  $^{3}\mathrm{He}^{*}$ and $^{4}\mathrm{He}^{*}$ atomic collisions},\ }\href
  {https://doi.org/10.1103/PhysRevA.104.033317} {\bibfield  {journal} {\bibinfo
   {journal} {Phys. Rev. A}\ }\textbf {\bibinfo {volume} {104}},\ \bibinfo
  {pages} {033317} (\bibinfo {year} {2021})}\BibitemShut {NoStop}%
\bibitem [{\citenamefont {Carr}\ and\ \citenamefont
  {Castin}(2004)}]{PhysRevA.69.043611}%
  \BibitemOpen
  \bibfield  {author} {\bibinfo {author} {\bibfnamefont {L.~D.}\ \bibnamefont
  {Carr}}\ and\ \bibinfo {author} {\bibfnamefont {Y.}~\bibnamefont {Castin}},\
  }\bibfield  {title} {\bibinfo {title} {Limits of sympathetic cooling of
  fermions: The role of heat capacity of the coolant},\ }\href
  {https://doi.org/10.1103/PhysRevA.69.043611} {\bibfield  {journal} {\bibinfo
  {journal} {Phys. Rev. A}\ }\textbf {\bibinfo {volume} {69}},\ \bibinfo
  {pages} {043611} (\bibinfo {year} {2004})}\BibitemShut {NoStop}%
\bibitem [{eva()}]{evap_note}%
  \BibitemOpen
  \href@noop {} {}\bibinfo {note} {Assuming the phase space density of the
  mixture and the efficiency of the evaporation cycle both remain constant for
  \(T\) greater than \(T_C\) the exact functional form is given by
  \(K=\frac{1}{f(\eta)} \left(\frac{S_f(\tau,N_f)}{k_b N_f} + 6\zeta(4) \left(
  \frac{\bar{\omega}_f}{\bar{\omega}_b}\right)^3 \tau \right)
  -\frac{N_b}{N_f}\) where \(K\) is a constant of motion, \(\tau\) is the
  reduced temperature, \(f(\eta)\) is the average energy of an boson with
  energy greater than or equal to \(\eta k_b T\), \(\frac{S_f(\tau,N_f)}{k_b
  N_f}\) is the reduced entropy of a fermi gas (which purely a function of
  \(\tau\)). The initial assumption essentially fixes \(K\) and \(\eta\), and
  is hence why we can apply the equation to a range of initial conditions and
  evaporation heights. The previous equation is only strictly valid for
  temperatures less than \(T_C\).}\BibitemShut {Stop}%
\bibitem [{\citenamefont {Metcalf}\ and\ \citenamefont {Van~der
  Straten}(1999)}]{metcalf1999laser}%
  \BibitemOpen
  \bibfield  {author} {\bibinfo {author} {\bibfnamefont {H.~J.}\ \bibnamefont
  {Metcalf}}\ and\ \bibinfo {author} {\bibfnamefont {P.}~\bibnamefont {Van~der
  Straten}},\ }\bibinfo {title} {Laser cooling and trapping}\ (\bibinfo
  {publisher} {Springer-Verlag},\ \bibinfo {address} {New York},\ \bibinfo
  {year} {1999})\ Chap.~\bibinfo {chapter} {4}, pp.\ \bibinfo {pages}
  {53--56}\BibitemShut {NoStop}%
\bibitem [{\citenamefont {Swansson}\ \emph {et~al.}(2004)\citenamefont
  {Swansson}, \citenamefont {Baldwin}, \citenamefont {Hoogerland},
  \citenamefont {Truscott},\ and\ \citenamefont {Buckman}}]{swansson2004high}%
  \BibitemOpen
  \bibfield  {author} {\bibinfo {author} {\bibfnamefont {J.}~\bibnamefont
  {Swansson}}, \bibinfo {author} {\bibfnamefont {K.}~\bibnamefont {Baldwin}},
  \bibinfo {author} {\bibfnamefont {M.}~\bibnamefont {Hoogerland}}, \bibinfo
  {author} {\bibfnamefont {A.}~\bibnamefont {Truscott}},\ and\ \bibinfo
  {author} {\bibfnamefont {S.}~\bibnamefont {Buckman}},\ }\bibfield  {title}
  {\bibinfo {title} {A high flux, liquid-helium cooled source of metastable
  rare gas atoms},\ }\href@noop {} {\bibfield  {journal} {\bibinfo  {journal}
  {Applied Physics B}\ }\textbf {\bibinfo {volume} {79}},\ \bibinfo {pages}
  {485} (\bibinfo {year} {2004})}\BibitemShut {NoStop}%
\bibitem [{\citenamefont {Swansson}\ \emph {et~al.}(2007)\citenamefont
  {Swansson}, \citenamefont {Dall},\ and\ \citenamefont {Truscott}}]{LVIS}%
  \BibitemOpen
  \bibfield  {author} {\bibinfo {author} {\bibfnamefont {J.}~\bibnamefont
  {Swansson}}, \bibinfo {author} {\bibfnamefont {R.}~\bibnamefont {Dall}},\
  and\ \bibinfo {author} {\bibfnamefont {A.}~\bibnamefont {Truscott}},\
  }\bibfield  {title} {\bibinfo {title} {An intense cold beam of metastable
  helium},\ }\href {https://doi.org/10.1007/s00340-006-2472-y} {\bibfield
  {journal} {\bibinfo  {journal} {Applied Physics B}\ }\textbf {\bibinfo
  {volume} {86}},\ \bibinfo {pages} {485} (\bibinfo {year} {2007})}\BibitemShut
  {NoStop}%
\bibitem [{\citenamefont {Henson}\ \emph
  {et~al.}(2022{\natexlab{b}})\citenamefont {Henson}, \citenamefont {Thomas},
  \citenamefont {Mehdi}, \citenamefont {Burnett}, \citenamefont {Ross},
  \citenamefont {Hodgman},\ and\ \citenamefont {Truscott}}]{Henson:22}%
  \BibitemOpen
  \bibfield  {author} {\bibinfo {author} {\bibfnamefont {B.~M.}\ \bibnamefont
  {Henson}}, \bibinfo {author} {\bibfnamefont {K.~F.}\ \bibnamefont {Thomas}},
  \bibinfo {author} {\bibfnamefont {Z.}~\bibnamefont {Mehdi}}, \bibinfo
  {author} {\bibfnamefont {T.~G.}\ \bibnamefont {Burnett}}, \bibinfo {author}
  {\bibfnamefont {J.~A.}\ \bibnamefont {Ross}}, \bibinfo {author}
  {\bibfnamefont {S.~S.}\ \bibnamefont {Hodgman}},\ and\ \bibinfo {author}
  {\bibfnamefont {A.~G.}\ \bibnamefont {Truscott}},\ }\bibfield  {title}
  {\bibinfo {title} {Trap frequency measurement with a pulsed atom laser},\
  }\href {https://doi.org/10.1364/OE.455009} {\bibfield  {journal} {\bibinfo
  {journal} {Opt. Express}\ }\textbf {\bibinfo {volume} {30}},\ \bibinfo
  {pages} {13252} (\bibinfo {year} {2022}{\natexlab{b}})}\BibitemShut {NoStop}%
\bibitem [{\citenamefont {Hilborn}(1982)}]{doi:10.1119/1.12937}%
  \BibitemOpen
  \bibfield  {author} {\bibinfo {author} {\bibfnamefont {R.~C.}\ \bibnamefont
  {Hilborn}},\ }\bibfield  {title} {\bibinfo {title} {Einstein coefficients,
  cross sections, f values, dipole moments, and all that},\ }\href
  {https://doi.org/10.1119/1.12937} {\bibfield  {journal} {\bibinfo  {journal}
  {American Journal of Physics}\ }\textbf {\bibinfo {volume} {50}},\ \bibinfo
  {pages} {982} (\bibinfo {year} {1982})}\BibitemShut {NoStop}%
\bibitem [{\citenamefont {Shin}\ \emph {et~al.}(2016)\citenamefont {Shin},
  \citenamefont {Henson}, \citenamefont {Khakimov}, \citenamefont {Ross},
  \citenamefont {Dedman}, \citenamefont {Hodgman}, \citenamefont {Baldwin},\
  and\ \citenamefont {Truscott}}]{ECL}%
  \BibitemOpen
  \bibfield  {author} {\bibinfo {author} {\bibfnamefont {D.~K.}\ \bibnamefont
  {Shin}}, \bibinfo {author} {\bibfnamefont {B.~M.}\ \bibnamefont {Henson}},
  \bibinfo {author} {\bibfnamefont {R.~I.}\ \bibnamefont {Khakimov}}, \bibinfo
  {author} {\bibfnamefont {J.~A.}\ \bibnamefont {Ross}}, \bibinfo {author}
  {\bibfnamefont {C.~J.}\ \bibnamefont {Dedman}}, \bibinfo {author}
  {\bibfnamefont {S.~S.}\ \bibnamefont {Hodgman}}, \bibinfo {author}
  {\bibfnamefont {K.~G.~H.}\ \bibnamefont {Baldwin}},\ and\ \bibinfo {author}
  {\bibfnamefont {A.~G.}\ \bibnamefont {Truscott}},\ }\bibfield  {title}
  {\bibinfo {title} {Widely tunable, narrow linewidth external-cavity gain chip
  laser for spectroscopy between 1.0 - 1.1 $\mu$m},\ }\href
  {https://doi.org/10.1364/OE.24.027403} {\bibfield  {journal} {\bibinfo
  {journal} {Opt. Express}\ }\textbf {\bibinfo {volume} {24}},\ \bibinfo
  {pages} {27403} (\bibinfo {year} {2016})}\BibitemShut {NoStop}%
\bibitem [{\citenamefont {Peng}\ \emph {et~al.}(2014)\citenamefont {Peng},
  \citenamefont {Zhou}, \citenamefont {Long}, \citenamefont {Wang},\ and\
  \citenamefont {Zhan}}]{Peng:14}%
  \BibitemOpen
  \bibfield  {author} {\bibinfo {author} {\bibfnamefont {W.}~\bibnamefont
  {Peng}}, \bibinfo {author} {\bibfnamefont {L.}~\bibnamefont {Zhou}}, \bibinfo
  {author} {\bibfnamefont {S.}~\bibnamefont {Long}}, \bibinfo {author}
  {\bibfnamefont {J.}~\bibnamefont {Wang}},\ and\ \bibinfo {author}
  {\bibfnamefont {M.}~\bibnamefont {Zhan}},\ }\bibfield  {title} {\bibinfo
  {title} {Locking laser frequency of up to 40 {G}hz offset to a reference with
  a 10 {G}hz electro-optic modulator},\ }\href
  {https://doi.org/10.1364/OL.39.002998} {\bibfield  {journal} {\bibinfo
  {journal} {Opt. Lett.}\ }\textbf {\bibinfo {volume} {39}},\ \bibinfo {pages}
  {2998} (\bibinfo {year} {2014})}\BibitemShut {NoStop}%
\bibitem [{\citenamefont {Harada}\ \emph {et~al.}(2016)\citenamefont {Harada},
  \citenamefont {Aoki}, \citenamefont {Ezure}, \citenamefont {Kato},
  \citenamefont {Hayamizu}, \citenamefont {Kawamura}, \citenamefont {Inoue},
  \citenamefont {Arikawa}, \citenamefont {Ishikawa}, \citenamefont {Aoki},
  \citenamefont {Uchiyama}, \citenamefont {Sakamoto}, \citenamefont {Ito},
  \citenamefont {Itoh}, \citenamefont {Ando}, \citenamefont {Hatakeyama},
  \citenamefont {Hatanaka}, \citenamefont {Imai}, \citenamefont {Murakami},
  \citenamefont {Nataraj}, \citenamefont {Shimizu}, \citenamefont {Sato},
  \citenamefont {Wakasa}, \citenamefont {Yoshida},\ and\ \citenamefont
  {Sakemi}}]{Harada:16}%
  \BibitemOpen
  \bibfield  {author} {\bibinfo {author} {\bibfnamefont {K.}~\bibnamefont
  {Harada}}, \bibinfo {author} {\bibfnamefont {T.}~\bibnamefont {Aoki}},
  \bibinfo {author} {\bibfnamefont {S.}~\bibnamefont {Ezure}}, \bibinfo
  {author} {\bibfnamefont {K.}~\bibnamefont {Kato}}, \bibinfo {author}
  {\bibfnamefont {T.}~\bibnamefont {Hayamizu}}, \bibinfo {author}
  {\bibfnamefont {H.}~\bibnamefont {Kawamura}}, \bibinfo {author}
  {\bibfnamefont {T.}~\bibnamefont {Inoue}}, \bibinfo {author} {\bibfnamefont
  {H.}~\bibnamefont {Arikawa}}, \bibinfo {author} {\bibfnamefont
  {T.}~\bibnamefont {Ishikawa}}, \bibinfo {author} {\bibfnamefont
  {T.}~\bibnamefont {Aoki}}, \bibinfo {author} {\bibfnamefont {A.}~\bibnamefont
  {Uchiyama}}, \bibinfo {author} {\bibfnamefont {K.}~\bibnamefont {Sakamoto}},
  \bibinfo {author} {\bibfnamefont {S.}~\bibnamefont {Ito}}, \bibinfo {author}
  {\bibfnamefont {M.}~\bibnamefont {Itoh}}, \bibinfo {author} {\bibfnamefont
  {S.}~\bibnamefont {Ando}}, \bibinfo {author} {\bibfnamefont {A.}~\bibnamefont
  {Hatakeyama}}, \bibinfo {author} {\bibfnamefont {K.}~\bibnamefont
  {Hatanaka}}, \bibinfo {author} {\bibfnamefont {K.}~\bibnamefont {Imai}},
  \bibinfo {author} {\bibfnamefont {T.}~\bibnamefont {Murakami}}, \bibinfo
  {author} {\bibfnamefont {H.~S.}\ \bibnamefont {Nataraj}}, \bibinfo {author}
  {\bibfnamefont {Y.}~\bibnamefont {Shimizu}}, \bibinfo {author} {\bibfnamefont
  {T.}~\bibnamefont {Sato}}, \bibinfo {author} {\bibfnamefont {T.}~\bibnamefont
  {Wakasa}}, \bibinfo {author} {\bibfnamefont {H.~P.}\ \bibnamefont
  {Yoshida}},\ and\ \bibinfo {author} {\bibfnamefont {Y.}~\bibnamefont
  {Sakemi}},\ }\bibfield  {title} {\bibinfo {title} {Laser frequency locking
  with 46 {G}hz offset using an electro-optic modulator for magneto-optical
  trapping of francium atoms},\ }\href {https://doi.org/10.1364/AO.55.001164}
  {\bibfield  {journal} {\bibinfo  {journal} {Appl. Opt.}\ }\textbf {\bibinfo
  {volume} {55}},\ \bibinfo {pages} {1164} (\bibinfo {year}
  {2016})}\BibitemShut {NoStop}%
\bibitem [{\citenamefont {Thompson}\ and\ \citenamefont
  {Scholten}(2012)}]{ThompsonandScholten}%
  \BibitemOpen
  \bibfield  {author} {\bibinfo {author} {\bibfnamefont {D.~J.}\ \bibnamefont
  {Thompson}}\ and\ \bibinfo {author} {\bibfnamefont {R.~E.}\ \bibnamefont
  {Scholten}},\ }\bibfield  {title} {\bibinfo {title} {Narrow linewidth tunable
  external cavity diode laser using wide bandwidth filter},\ }\href@noop {}
  {\bibfield  {journal} {\bibinfo  {journal} {Review of Scientific
  Instruments}\ }\textbf {\bibinfo {volume} {83}},\ \bibinfo {pages} {023107}
  (\bibinfo {year} {2012})}\BibitemShut {NoStop}%
\bibitem [{\citenamefont {Okoshi}\ \emph {et~al.}(1980)\citenamefont {Okoshi},
  \citenamefont {Kikuchi},\ and\ \citenamefont {Nakayama}}]{linewidth}%
  \BibitemOpen
  \bibfield  {author} {\bibinfo {author} {\bibfnamefont {T.}~\bibnamefont
  {Okoshi}}, \bibinfo {author} {\bibfnamefont {K.}~\bibnamefont {Kikuchi}},\
  and\ \bibinfo {author} {\bibfnamefont {A.}~\bibnamefont {Nakayama}},\
  }\bibfield  {title} {\bibinfo {title} {Novel method for high resolution
  measurement of laser output spectrum},\ }\href@noop {} {\bibfield  {journal}
  {\bibinfo  {journal} {Electron. Lett.}\ }\textbf {\bibinfo {volume} {16}},\
  \bibinfo {pages} {630 – 631} (\bibinfo {year} {1980})}\BibitemShut
  {NoStop}%
\bibitem [{\citenamefont {Manning}\ \emph {et~al.}(2013)\citenamefont
  {Manning}, \citenamefont {RuGway}, \citenamefont {Hodgman}, \citenamefont
  {Dall}, \citenamefont {Baldwin},\ and\ \citenamefont
  {Truscott}}]{Manning_2013}%
  \BibitemOpen
  \bibfield  {author} {\bibinfo {author} {\bibfnamefont {A.~G.}\ \bibnamefont
  {Manning}}, \bibinfo {author} {\bibfnamefont {W.}~\bibnamefont {RuGway}},
  \bibinfo {author} {\bibfnamefont {S.~S.}\ \bibnamefont {Hodgman}}, \bibinfo
  {author} {\bibfnamefont {R.~G.}\ \bibnamefont {Dall}}, \bibinfo {author}
  {\bibfnamefont {K.~G.~H.}\ \bibnamefont {Baldwin}},\ and\ \bibinfo {author}
  {\bibfnamefont {A.~G.}\ \bibnamefont {Truscott}},\ }\bibfield  {title}
  {\bibinfo {title} {Third-order spatial correlations for ultracold atoms},\
  }\href {https://doi.org/10.1088/1367-2630/15/1/013042} {\bibfield  {journal}
  {\bibinfo  {journal} {New Journal of Physics}\ }\textbf {\bibinfo {volume}
  {15}},\ \bibinfo {pages} {013042} (\bibinfo {year} {2013})}\BibitemShut
  {NoStop}%
\bibitem [{\citenamefont {Dall}\ \emph {et~al.}(2013)\citenamefont {Dall},
  \citenamefont {Manning}, \citenamefont {Hodgman}, \citenamefont {RuGway},
  \citenamefont {Kheruntsyan},\ and\ \citenamefont {Truscott}}]{Dall2013}%
  \BibitemOpen
  \bibfield  {author} {\bibinfo {author} {\bibfnamefont {R.~G.}\ \bibnamefont
  {Dall}}, \bibinfo {author} {\bibfnamefont {A.~G.}\ \bibnamefont {Manning}},
  \bibinfo {author} {\bibfnamefont {S.~S.}\ \bibnamefont {Hodgman}}, \bibinfo
  {author} {\bibfnamefont {W.}~\bibnamefont {RuGway}}, \bibinfo {author}
  {\bibfnamefont {K.~V.}\ \bibnamefont {Kheruntsyan}},\ and\ \bibinfo {author}
  {\bibfnamefont {A.~G.}\ \bibnamefont {Truscott}},\ }\bibfield  {title}
  {\bibinfo {title} {Ideal n-body correlations with massive particles},\ }\href
  {http://dx.doi.org/10.1038/nphys2632} {\bibfield  {journal} {\bibinfo
  {journal} {Nature Physics}\ }\textbf {\bibinfo {volume} {9}},\ \bibinfo
  {pages} {341} (\bibinfo {year} {2013})}\BibitemShut {NoStop}%
\bibitem [{\citenamefont {Hodgman}\ \emph {et~al.}(2017)\citenamefont
  {Hodgman}, \citenamefont {Khakimov}, \citenamefont {Lewis-Swan},
  \citenamefont {Truscott},\ and\ \citenamefont
  {Kheruntsyan}}]{PhysRevLett.118.240402}%
  \BibitemOpen
  \bibfield  {author} {\bibinfo {author} {\bibfnamefont {S.~S.}\ \bibnamefont
  {Hodgman}}, \bibinfo {author} {\bibfnamefont {R.~I.}\ \bibnamefont
  {Khakimov}}, \bibinfo {author} {\bibfnamefont {R.~J.}\ \bibnamefont
  {Lewis-Swan}}, \bibinfo {author} {\bibfnamefont {A.~G.}\ \bibnamefont
  {Truscott}},\ and\ \bibinfo {author} {\bibfnamefont {K.~V.}\ \bibnamefont
  {Kheruntsyan}},\ }\bibfield  {title} {\bibinfo {title} {Solving the quantum
  many-body problem via correlations measured with a momentum microscope},\
  }\href {https://doi.org/10.1103/PhysRevLett.118.240402} {\bibfield  {journal}
  {\bibinfo  {journal} {Phys. Rev. Lett.}\ }\textbf {\bibinfo {volume} {118}},\
  \bibinfo {pages} {240402} (\bibinfo {year} {2017})}\BibitemShut {NoStop}%
\bibitem [{\citenamefont {Cowan}(2019)}]{cowan2019chemical}%
  \BibitemOpen
  \bibfield  {author} {\bibinfo {author} {\bibfnamefont {B.}~\bibnamefont
  {Cowan}},\ }\bibfield  {title} {\bibinfo {title} {On the chemical potential
  of ideal fermi and bose gases},\ }\href@noop {} {\bibfield  {journal}
  {\bibinfo  {journal} {Journal of Low Temperature Physics}\ }\textbf {\bibinfo
  {volume} {197}},\ \bibinfo {pages} {412} (\bibinfo {year}
  {2019})}\BibitemShut {NoStop}%
\end{thebibliography}%

\end{document}